# Pressure-Induced Structural and Magnetic Evolution in Layered Antiferromagnet YbMn$_2$Sb$_2$

*Mingyu Xu[1], Matt Boswell[1,2], Aya Rutherford[3], Cheng Peng[1], Ying Zhou[4], Shuyang Wang[5], Zhaorong Yang[4,5], Antonio M. dos Santos[2], Haidong Zhou[3], Weiwei Xie[1*]*

1. Department of Chemistry, Michigan State University, East Lansing, MI, 48864, USA
2. Neutron Scattering Division, Oak Ridge National Laboratory, Oak Ridge, TN 37831 USA
3. Department of Physics and Astronomy, University of Tennessee, Knoxville, TN 37996, USA
4. Institutes of Physical Science and Information Technology, Anhui University, Hefei 230601, China
5. Anhui Province Key Laboratory of Low-Energy Quantum Materials and Devices, High Magnetic Field Laboratory, HFIPS, Chinese Academy of Sciences, Hefei, Anhui 230031, China

[*]Corresponding author: Dr. Weiwei Xie (xieweiwe@msu.edu)

*Abstract*

Electronic states under pressure exhibit unconventional spin and charge dynamics that provide a powerful route to uncover exotic phases in quantum materials. Here, we present the structural, magnetic, and electronic evolution of YbMn$_2$Sb$_2$ under pressure. Single-crystal X-ray diffraction reveals a pressure-induced structural transition from the space group trigonal $P\bar{3}m1$ to the monoclinic $P2_1/m$ phase near 3.5 GPa, which remains stable up to 10 GPa. Magnetization measurements display an anomalously weak net magnetic moment and the absence of Curie-Weiss behavior up to 400 K, suggesting the formation of short-range Mn moment pairs that cancel macroscopically and subsequently evolve into long-range order upon cooling. Temperature-dependent resistivity shows semiconducting behavior with a transition at ~119 K at ambient pressure, while pressure induces a dramatic suppression of resistance and the emergence of metallic-like temperature dependence, stabilized beyond 5 GPa. This pressure-driven semiconductor-metal transition is consistent with our density functional theory calculations, confirming the closing of the band gap under compression. Neutron diffraction under pressure identifies an incommensurate magnetic structure with antiparallel correlations between paired spins. Together, these results demonstrate how pressure-driven structural tuning and competing exchange interactions stabilize unconventional magnetic states in this low-dimensional magnetic semiconductor.

# Introduction

Understanding the relation between magnetic properties and structure is central to advancing the study of quantum materials and underpins the design of next-generation technologies.[1–3] High-pressure techniques offer a particularly powerful experimental route, as they simultaneously tune crystal structures and electronic bands, thereby exposing the fundamental interplay between structural perturbation and magnetic behavior. Consequently, high pressure serves as an essential tool for probing the coupled magnetic and structural responses that define the emergent properties of quantum materials. The $AM_2X_2$ family of compounds, where $A$ is an alkali, alkaline-earth, or rare-earth element, $M$ is a transition metal, and $X$ is a pnictogen or chalcogen, has attracted significant attention over the past decades due to their diverse physical properties, including magnetism[4–7], topology[8–10], thermoelectricity[11–13], and superconductivity[14–17]. These materials typically crystallize in either the tetragonal $ThCr_2Si_2$ structure or the trigonal $CaAl_2Si_2$ structure. In the latter, the $MX$ layers form a corrugated honeycomb substructure, a structure well known for hosting competing magnetic interactions[18]. This makes the $REM_2X_2$ ($RE$ is rare earth) family an ideal platform for studying magnetic properties arising from $3d$ and $4f$ electrons, as well as topological phenomena driven by strong spin-orbit coupling from heavy pnictogen or chalcogen atoms.

In situ high-pressure measurements are an effective approach for probing quantum materials, and the study of $AM_2X_2$ compounds under compression offers valuable opportunities to elucidate their magnetic and topological properties while uncovering emergent phases. A representative example is $CaMn_2Bi_2$. The trigonal $La_2O_3$-type compound $Mg_3Bi_2$ has been established as an ideal platform for realizing type-II nodal-line topological electronic states[19]. Similarly, $CaMn_2Bi_2$ crystallizes in the same $La_2O_3$-type structure with the space group $P\bar{3}m1$ at ambient pressure, in which Mn-Bi layers are separated by Ca layers[20]. Under pressure between 2 and 3 GPa, $CaMn_2Bi_2$ undergoes a distinctive plane-to-chain structural transition accompanied by a large volume collapse[21]. Across this transition, the puckered Mn-Mn honeycomb layers convert into quasi-one-dimensional zigzag chains. At higher pressures, incommensurate spiral spin order emerges, driven by the interplay between Mn magnetism and strong spin-orbit coupling from Bi; sinusoidal spin order has been observed at pressures up to 7.4 GPa[22].

Similarly, YbMn$_2$Sb$_2$ crystallizes in the La$_2$O$_3$-type structure with the trigonal $P\bar{3}m1$ space group[23], with antiferromagnetic long-range order around 119 K[24–28]. Measurements on single-crystalline YbMn$_2$Sb$_2$ exhibit insulating behavior and the absence of long-range magnetic order to the lowest measured temperature, despite the presence of localized 4$f$ brought by Yb[29] and 3$d$ moments[30]. Instead, it stabilizes a static, strongly disordered antiferromagnetic ground state in which spin-lattice coupling plays a key role, a behavior distinct from that of ferromagnetic polycrystalline samples[28]. The combination of structural tunability and complex magnetism makes YbMn$_2$Sb$_2$ an excellent platform using pressure techniques for probing the interplay between structure and magnetism and for elucidating how 4$f$ and 3$d$ electrons cooperate to stabilize exotic magnetic order.

Motivated by the strong coupling between crystal structure, electronic states, and magnetism in layered Mn pnictides, we investigate how external pressure tunes the structural, electronic, and magnetic properties of YbMn$_2$Sb$_2$. Pressure provides a clean and continuous control parameter to modify interlayer spacing, bonding geometry, and exchange pathways without introducing chemical disorder. By combining temperature- and pressure-dependent diffraction experiments with electronic structure calculations and transport and magnetic measurements, this work aims to elucidate how structural instabilities and competing exchange interactions cooperate to stabilize emergent electronic and magnetic states in this family of materials.

## Experimental Methods

**Crystal Synthesis:** Single-crystalline YbMn$_2$Sb$_2$ was synthesized using a two-step high-temperature solution growth method[31]. In the first step, Yb (Thermo Scientific, 99.9% RE), Mn (BTC, 99.85% trace metal), Sb (Alfa Aesar, 99.5%), and Sn (Alfa Aesar, 99%) were combined in a molar ratio of 1:2:2:25 and loaded into fritted Canfield Crucible Sets (CCS)[32,33]. The mixture was heated to 1150 °C at a rate of 180 °C/hour and held at this temperature for 12 hours, followed by cooling to 700 °C over 24 hours. At 700 °C, the solid phases were separated from the remaining liquid using a laboratory centrifuge. High-pressure resistance measurements used the sample from the 700 °C separation. In the second step, the decanted liquid was resealed, reheated to 1150 °C at 180 °C/hour, and held for 12 hours. It was then rapidly cooled to 750 °C, followed by slow cooling to 320 °C at a rate of 4 °C/hour. At 320 °C, the melt was decanted, and single-crystalline YbMn$_2$Sb$_2$ was separated from the excess flux. Ambient pressure measurements and high-pressure single crystal X-ray diffraction measurements used the samples from the second step. Both samples were characterized using X-ray diffraction measurements.

**Ambient pressure Single Crystal X-ray diffraction (SCXRD):** A YbMn$_2$Sb$_2$ single crystal (0.11 × 0.07 × 0.02 mm³) was mounted on a nylon loop with Paratone oil and examined using a Rigaku XtalLAB Synergy, Dualflex, Hypix diffractometer. Measurements were performed between 100-400 K at ambient pressure. Data were collected using ω scans with Mo Kα radiation (λ = 0.71073 Å) from a micro-focus sealed tube (50 kV, 1 mA). Collection strategies, including the number of runs and images, were determined automatically by *CrysAlisPro* (v1.171.42.101a, Rigaku OD, 2023). Data reduction included Lorentz and polarization corrections, numerical absorption correction via Gaussian integration over a multifaceted crystal model[34], and an empirical spherical harmonics correction using the SCALE3 ABSPACK algorithm[35]. Structures were solved and refined with the Bruker SHELXTL package[36,37]. Results for the ambient and selected pressures are listed in **Tables 1** and **2.**

**High-Pressure SCXRD:** High-pressure single-crystal XRD was performed at room temperature on the same YbMn$_2$Sb$_2$ crystal characterized at ambient pressure, using a Diacell One20DAC (Almax-easyLab) with 500 μm culet Boehler-Almax anvils. A 250 μm-thick stainless-steel gasket was pre-indented to 44 μm, and a 210 μm hole was drilled by Advanced Electric Discharge Machine (Boehler μDriller, Almax easyLab) to hold the sample. A 4:1 methanol-ethanol mixture served as the pressure medium to maintain hydrostatic conditions[38]. Applied pressures (up to 8.8 GPa) were determined from the R$_1$ fluorescence line of ruby[39–41].

**Magnetization and resistance measurements:** Temperature- and magnetic-field-dependent magnetization and resistance measurements were performed using Quantum Design (QD) Magnetic Property Measurement System (MPMS3) and Physical Property Measurement System (PPMS), respectively. For magnetization measurements with *H* ∥ *a*, the plate-like single crystal was mounted between two collapsed plastic straws, with a third uncollapsed straw serving as an outer sheath or secured in a quartz sample holder. For *H* ∥ *c*, the sample was supported between two cut straws wrapped with Teflon tape, with a third uncollapsed straw used as an outer sheath. DC electrical resistance was measured using the DynaCool

PPMS in a standard four-contact geometry with a constant current of 0.1 mA. Platinum wires (50 μm diameter) were bonded to the sample using silver paint (DuPont 4929N). The magnetic field, up to 90 kOe, was applied along the *c*-axis, perpendicular to the current in the *ab* plane.

**High-pressure Electrical Resistance Measurements:** High-pressure electrical resistance measurements on single-crystalline YbMn$_2$Sb$_2$ were carried out on a commercial Physical Property Measurement System (PPMS). High pressure was generated in a Be-Cu diamond anvil cell equipped with 300 μm culet anvils. The T301 stainless-steel gasket, initially 200 μm thick, was pre-indented to 30 μm, and a 280 μm hole was drilled using an Advanced Electric Discharge Machine (Boehler μDriller, Almax easyLab) at its center. To electrically insulate the leads from the metallic gasket, the gasket surface was coated with a fine mixture of cubic boron nitride powder and epoxy. The crystal was positioned at the center of the gasket hole in a standard van der Pauw four-probe configuration, with thin Pt foils (0.004±0.0025 mm, Thermo Scientific) serving as electrodes. Two experimental runs were performed on separate single crystals with NaCl as the pressure-transmitting medium. Pressures were applied at room temperature and determined via the ruby fluorescence method[42].

**Electronic Structure Calculations:** The density functional theory (DFT) calculations using version 7.4.1 of the Quantum ESPRESSO code[43,44] were carried out to investigate the band structures and density of states (DOS) of YbMn$_2$Sb$_2$ structures refined under ambient and high pressure. The calculations were performed using SSSP Efficiency v1.3.0 pseudopotentials[45] with the Perdew-Burke-Ernzerhof (PBE) exchange-correlation functional[46]. A wavefunction cutoff energy of 100 Ry was used, with the charge density cutoff set to twelve times this value. An 11×11×6 Monkhorst-Pack *k*-points mesh[47] was employed in the reciprocal space. Convergence tests were conducted to ensure the change of total energy was smaller than 1 meV/atom. The Davidson diagonalization algorithm[48] was applied, and the convergence threshold for self-consistency is set to $10^{-9}$ Ry. The high symmetry path was generated with the Spglib library[49], following the convention of Setyawan and Curtarolo[50].

**High-pressure neutron diffraction measurements:** High-pressure neutron diffraction experiments were performed at the SNAP beamline (BL-3) at Oak Ridge National Laboratory (ORNL). Powdered YbMn$_2$Sb$_2$ (~500 mg each) was loaded without a pressure medium into a single-toroidal anvil. Pressure was generated using a Paris–Edinburgh press cooled with liquid nitrogen. The detector banks at 90° and 50° were used to collect nuclear and magnetic scattering data, respectively. Pressure of the hydraulic fluid was increased in ~75 bar steps (~0.5 GPa) up to 1200 bar (~8.2 GPa), with temperature-dependent measurements taken at selected pressure points. Lead was included with the YbMn$_2$Sb$_2$ sample for in situ pressure calibration[51]. Nuclear diffraction patterns from the 90° bank were refined using GSAS-II[52], while magnetic scattering data from the 50° bank were analyzed with the FullProf suite[53]. Magnetic space groups and irreducible representations were determined using the SARAh software package[54].

**Energy Dispersive Spectroscopy (EDS):** The sample phase composition was analyzed employing a JEOL 6610LV scanning electron microscope equipped with a tungsten hairpin emitter (JEOL Ltd., Tokyo, Japan). For elemental analysis, energy-dispersive X-ray spectroscopy was conducted utilizing an Oxford Instruments AZtec system (Oxford Instruments, High Wycomb, Buckinghamshire, England), operating software version 3.1. This setup included a 20 mm² Silicon Drift Detector (SDD) and an ultra-thin window integrated with the JEOL 6610LV SEM. The plate-like crystals were affixed to carbon adhesive tape and introduced into the SEM chamber for examination at an accelerating voltage of 20 kV. Data acquisition entailed collecting spectra at multiple points along the individual crystals over an optimized timeframe. Semi-quantitative compositional analysis was performed using SEM Quant software, which applies corrections for matrix effects to the intensity measurements. The data is shown in Fig. S

## Results and Discussion

**Tables 1** and **2** summarize the crystal structure refinements and atomic coordinates of YbMn$_2$Sb$_2$ collected at 100 K and 300 K under ambient pressure, as well as at room temperature under applied pressures of 3.2 GPa and 8.8 GPa (representative datasets are shown). The same crystal was used for all high-pressure measurements. Structural refinements confirmed the $P\bar{3}m1$ space group at ambient pressure and $P2_1/m$ space group under high pressure. For temperature-dependent SCXRD measurements, a larger crystal (0.19 × 0.112 × 0.1mm, 14 times larger than the high-pressure SCXRD sample) was selected, as multiple temperatures were required to evaluate potential unit cell changes rather than to refine atomic positions precisely. The relatively large $R_{int}$ value at the low temperature may instead arise from residual strain in the crystal. The refinements yielded goodness-of-fit values of approximately 1.1, except at pressures near the structural transition. The largest residual electron density peak and hole in the final difference Fourier map were both less than 10 e$^-$/Å$^3$.

**Table 1.** Crystal structure and refinement data for YbMn$_2$Sb$_2$ at 100 K and 300 K (ambient pressure) and at 300 K under 8.8 GPa and 3.2 GPa. Values in parentheses represent estimated standard deviations from the refinements.

| Temperature (K) | 300 | 100 | 300 | 300 |
|---|---|---|---|---|
| Pressure (GPa) | Ambient | Ambient | 3.2 | 8.8 |
| Space Group | $P\bar{3}m1$ | $P\bar{3}m1$ | $P\bar{3}m1$ | $P2_1/m$ |
| Unit Cell dimensions | $a$ = 4.5269(5) Å; $b$ = 4.5269(5) Å; $c$ = 7.4570(9) Å | $a$ = 4.5214(2) Å; $b$ = 4.5214(2) Å; $c$ = 7.4157(3) Å | $a$ = 4.4326(5) Å; $b$ = 4.4326(5) Å; $c$ = 7.2716(9) Å | $a$ = 7.003(2) Å; $b$ = 3.9742(11) Å; $c$ = 7.270(2) Å; $\beta$ = 94.22(3) |
| Volume | 132.34(2) Å$^3$ | 131.292(9) | 123.73(3) | 201.79(11) |
| Z | 1 | 1 | 1 | 2 |
| Density (calculated) | 6.61 | 6.66 | 7.07 | 8.66 |
| Absorption coefficient | 32.0 | 32.2 | 34.3 | 21.0 |
| F (000) | 222.0 | 221.6 | 222.0 | 222.0 |
| 2θ range | 10.402 to 82.538 | 5.5 to 82.46 | 10.624 to 74.886 | 11.682 to 74.868 |
| Reflections collected | 3759 | 7669 | 1507 | 2649 |
| Independent reflections | 371 [$R_{int}$ = 0.0979] | 374 [$R_{int}$ = 0.1518] | 269 [$R_{int}$ = 0.1026] | 623 [$R_{int}$ = 0.3230] |
| Data/restraints/parameters | 371/0/9 | 374/0/10 | 269/0/10 | 623/0/31 |
| Final $R$ indices | $R_1$ (I>2σ(I)) = 0.0278; $wR_2$ (I > 2σ(I)) = 0.0617; $R_1$ (all) = 0.0339; $wR_2$ (all) = 0.0638 | $R_1$ (I>2σ(I)) = 0.0285; $wR_2$ (I > 2σ(I)) = 0.0621; $R_1$ (all) = 0.0457; $wR_2$ (all) = 0.0851 | $R_1$ (I>2σ(I)) = 0.0655; $wR_2$ (I > 2σ(I)) = 0.1911; $R_1$ (all) = 0.1180; $wR_2$ (all) = 0.3461 | $R_1$ (I>2σ(I)) = 0.1960; $wR_2$ (I > 2σ(I)) = 0.3936; $R_1$ (all) = 0.3491; $wR_2$ (all) = 0.4880 |
| Largest diff. peak and hole | +3.10 e$^-$/Å$^3$ and -2.44 e$^-$/Å$^3$ | +7.83 e$^-$/Å$^3$ and -11.30 e$^-$/Å$^3$ | +9.64 e$^-$/Å$^3$ and -10.00 e$^-$/Å$^3$ | +4.73 e$^-$/Å$^3$ and -5.68 e$^-$/Å$^3$ |
| Goodness-of-fit on F$^2$ | 1.046 | 1.091 | 1.372 | 1.184 |

**Table 2.** Atomic coordinates and equivalent isotropic atomic displacement parameters (Å$^2$) for YbMn$_2$Sb$_2$ at 100 K and 300 K (ambient pressure) and at 300 K under 8.8 GPa and 3.2 GPa.

| P (GPa) | T (K) | Atom | Wyck. | x | y | z | Occ. | U$_{eq}$ |
|---|---|---|---|---|---|---|---|---|
| Ambient | 300 | Yb | 1a | 0 | 0 | 0 | 1 | 0.014(12) |
| | | Sb | 2d | 1/3 | 2/3 | 0.25061(7) | 1 | 0.012(11) |
| | | Mn | 2d | 1/3 | 2/3 | 0.62348(17) | 1 | 0.015(9) |
| Ambient | 100 | Yb | 1a | 0 | 0 | 0 | 1 | 0.008(19) |
| | | Sb | 2d | 1/3 | 2/3 | 0.25029(9) | 1 | 0.008(18) |
| | | Mn | 2d | 1/3 | 2/3 | 0.62230(2) | 1 | 0.009(9) |
| 3.2 | 300 | Yb | 1a | 0 | 0 | 0 | 1 | 0.017(9) |
| | | Sb | 2d | 1/3 | 2/3 | 0.24750(3) | 1 | 0.015(9) |
| | | Mn | 2d | 1/3 | 2/3 | 0.62230(9) | 1 | 0.021(17) |
| 8.8 | 300 | Yb | 2i | 0.51860(6) | 1/4 | 0.75280(11) | 1 | 0.050(3) |
| | | Sb$_1$ | 2i | 0.23820(10) | 1/4 | 0.42130(19) | 1 | 0.064(5) |
| | | Sb$_2$ | 2i | 0.75470(9) | 1/4 | 0.11940(18) | 1 | 0.046(4) |
| | | Mn$_1$ | 2i | 0.00300(20) | 1/4 | 0.70700(50) | 1 | 0.085(13) |
| | | Mn$_2$ | 2i | 0.12100(20) | 1/4 | 0.06400(50) | 1 | 0.077(14) |

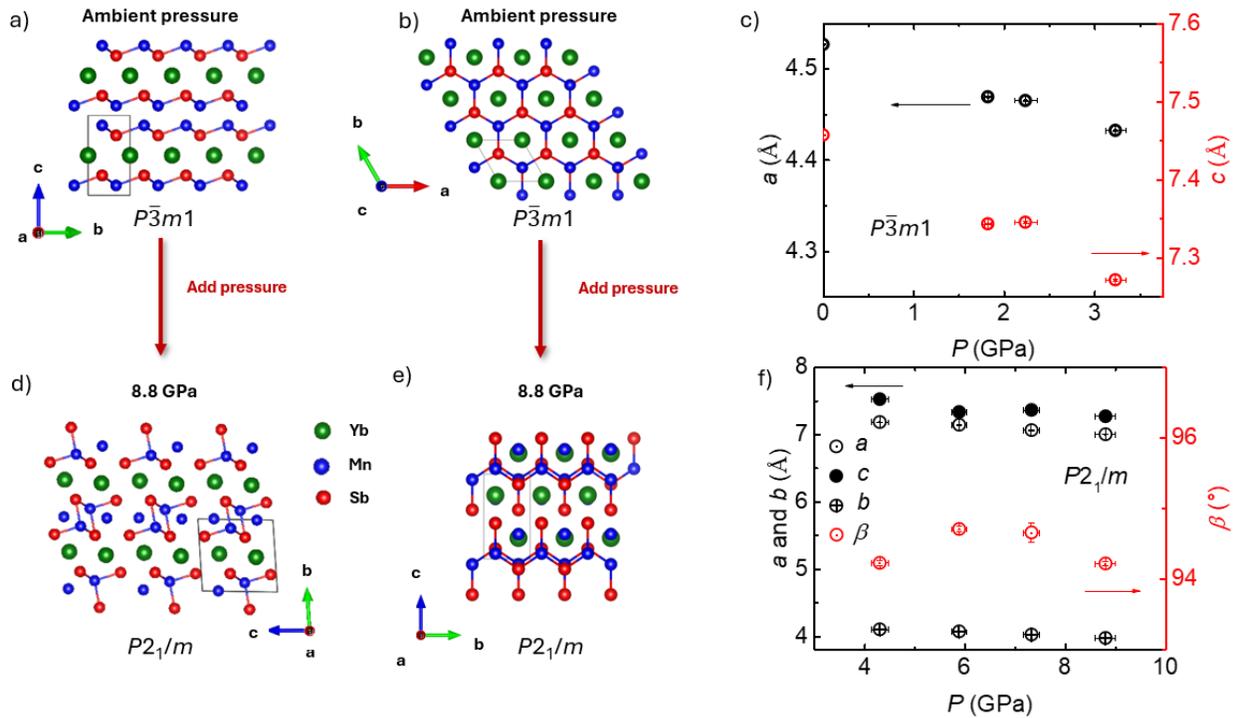

**Figure 1. Crystal Structure of YbMn$_2$Sb$_2$ and pressure-dependent single crystal X-ray diffraction results**. Panels (*a*) and (*b*) illustrate the ambient-pressure $P\bar{3}m1$ structure, while panels (*d*) and (*e*) depict the high-pressure $P2_1/m$ structure generated by VESTA software.[52] Panels (*c*) and (*f*) present the evolution of lattice parameters as a function of applied pressure.

YbMn$_2$Sb$_2$ crystallizes in a La$_2$O$_3$-type structure with the trigonal $P\bar{3}m1$ space group[28]. As shown in **Figures 1***a* and **1***b*, at ambient pressure, Yb, Mn, and Sb occupy distinct atomic layers stacked along the *c*-axis. Upon compression, both the *a* and *c* lattice parameters generally decrease (**Figure 1***c*). At 8.8 GPa (**Figures 1***d* and **1***e*), the structure adopts a denser atomic packing in which the distinct layered character is lost. The in-plane geometry deviates from the ideal hexagonal arrangement, and the Mn sublattice undergoes a transformation from a puckered honeycomb network to parallel one-dimensional zigzag ladder-like chains. In this monoclinic phase, the Mn-Mn separation along the chain is 3.197(2) Å, while the nearest-neighbor Mn-Mn distances are considerably shorter, ranging from 2.62(3) Å to 2.84(3) Å. This pressure-induced transition closely resembles that observed in CaMn$_2$Bi$_2$ near 3 GPa[21]. The detailed structural refinements are summarized in **Tables 1** and **2**, based on high-pressure single-crystal X-ray diffraction measurements. At the highest pressure, diffraction intensities became noticeably weaker and data completeness at high angles decreased, consistent with increased non-hydrostatic strain in the DAC. Residual density peaks/holes are located close to existing heavy-atom sites (Yb) and neighboring atoms and are attributed to Fourier truncation, limited resolution/completeness, and residual absorption effects under DAC conditions. No chemically meaningful disorder or missing atoms are indicated.

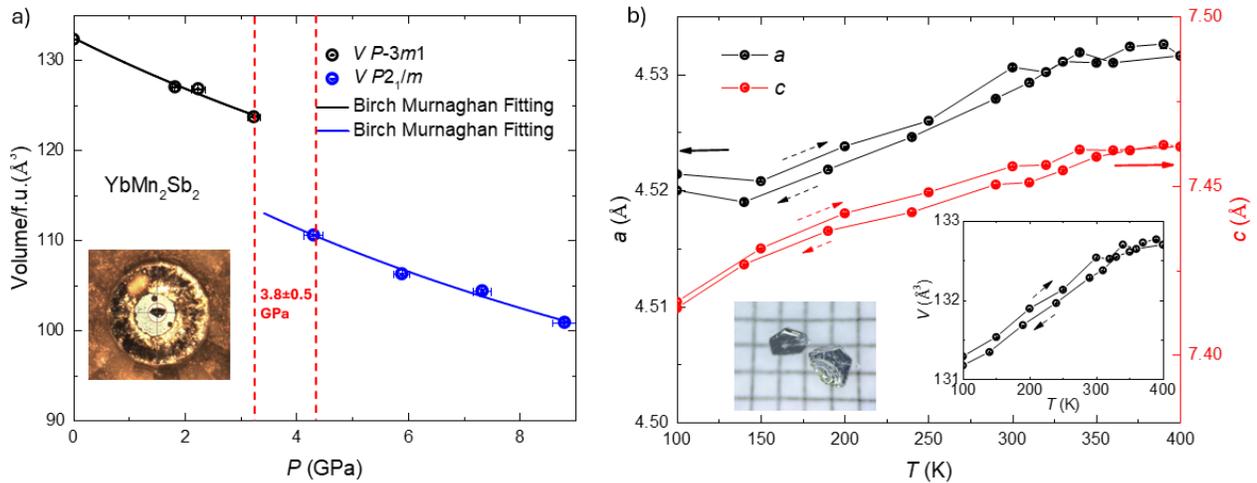

**Figure 2. Pressure- and temperature-dependent evolution of the unit-cell volume and lattice parameters of YbMn$_2$Sb$_2$.** (*a*) Unit-cell volume as a function of pressure with a Birch-Murnaghan equation-of-state fit. The inset shows the gasket and sample chamber at 8.8 GPa. (*b*) Temperature dependence of the lattice parameters *a* and *c* between 100 and 400 K. The right inset displays the corresponding unit-cell volume change with temperature, while the left inset shows an optical image of synthesized YbMn$_2$Sb$_2$ single crystals on millimeter grid paper. Dashed arrows indicate the direction of temperature variation.

**Figure 2a** shows the volume change when pressure is applied to the single crystal YbMn$_2$Sb$_2$, along with the Birch-Murnaghan fit. Given the Limited number of pressure points and small pressure ranges, we fixed the pressure derivative of bulk modulus $B_0'$ to 4, corresponding to a second-order Birch-Murnaghan equation. The trigonal phase $B_{0\text{-trigonal}}$ = 42.17 GPa and monoclinic phase $B_{0\text{-monoclinic}}$ = 26.17 GPa, indicating an increase in compressibility across the phase transition. Such behavior can arise when pressure-induced structural rearrangement introduces additional lattice degrees of freedom and enhances elastic anisotropy, leading to a more compliant volumetric response[55]. At approximately 3.8 ± 0.5 GPa, a discontinuity appears in the Birch-Murnaghan fit, accompanied by a space group change from trigonal $P\bar{3}m1$ to monoclinic $P2_1/m$, indicating a structural phase transition (**Figure 2a**). The inset of **Figure 2a** shows the sample mounted in a diamond anvil cell for single-crystal X-ray diffraction; the two points are ruby spheres used for pressure calibration, and the red circle represents a 0.1 mm scale. Above 4 GPa, the structure shows the monoclinic $P2_1/m$ symmetry, with both lattice parameters and unit-cell volume continuing to decrease with pressure (**Figure 1f** and **Figure 2a**). **Figure 2b** shows the temperature dependence of the lattice parameters of YbMn$_2$Sb$_2$ between 100 K and 400 K to probe possible structural changes. The data is collected during temperature increases and then decreases, shown using dashed arrows. The inset displays the unit-cell volume during heating and cooling, revealing no evidence of a transition. The volume increases monotonically with temperature within our accessible temperature range.

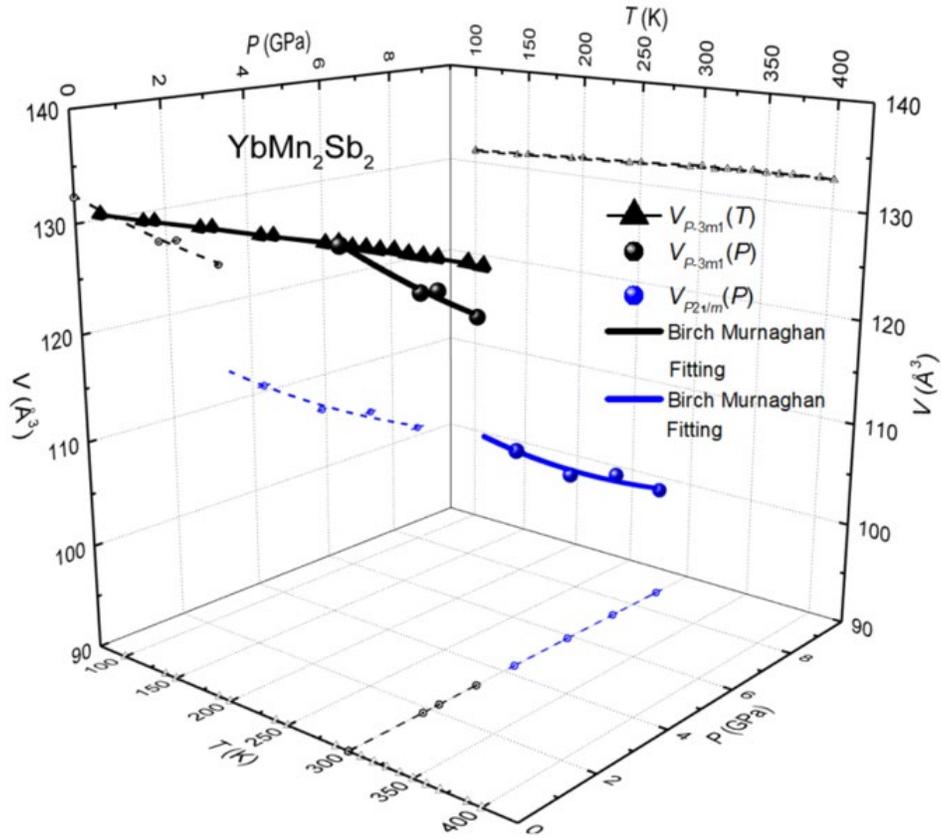

**Figure 3. Summary of the unit-cell volume evolution of YbMn$_2$Sb$_2$ as a function of temperature and pressure.** Projected data are represented by hollow spheres connected with a dashed line.

**Figure 3** presents a three-dimensional representation of the unit-cell volume of YbMn$_2$Sb$_2$ as a function of both temperature and pressure. While thermal expansion between 100 K and 400 K produces a measurable increase of about 1% in volume, the effect is significantly smaller than the reduction induced by applied pressure (23% volume change). The absence of anomalies in this temperature range confirms that no structural transition occurs at ambient pressure. Analysis of the pressure-volume relationship further reveals that a suppression of approximately 10% in unit-cell volume is required to drive the structural phase transition in YbMn$_2$Sb$_2$.

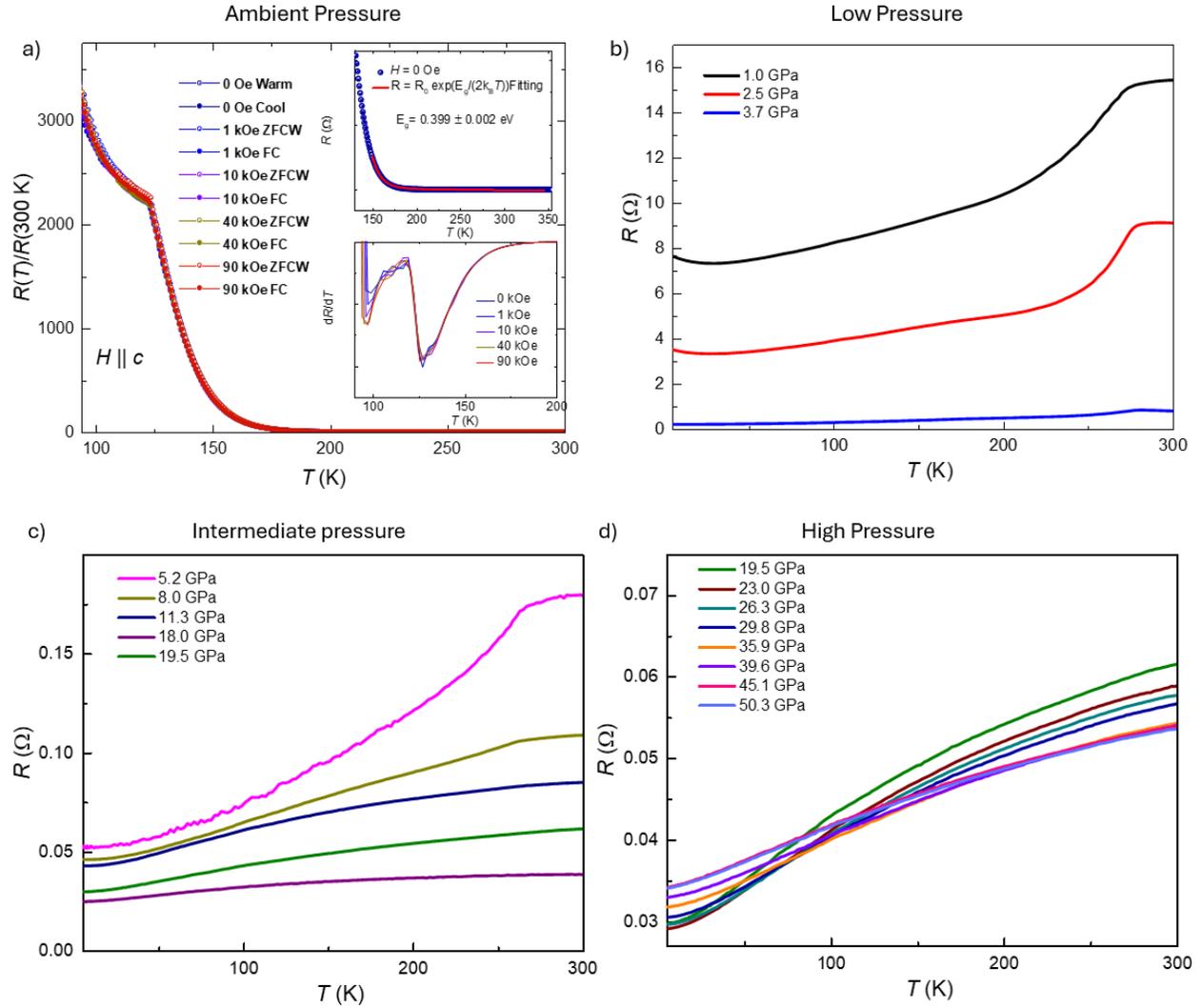

**Figure 4. Temperature-dependent resistance measurements at ambient conditions with different applied fields and high pressure with zero applied field.** (*a*) Resistance as a function of temperature measured under different magnetic fields. The top inset shows the resistance fitting at zero field, and the bottom inset presents the derivative of resistance. (*b*) Resistance versus temperature at pressures between 1 and 3.7 GPa. (*c*) Resistance versus temperature at intermediate pressures between 5.2 and 19.5 GPa. (*d*) Resistance versus temperature at higher pressures between 19.5 and 50.3 GPa.

**Figure 4a** shows the normalized resistance, $R(T)/R(300K)$, as a function of temperature under various magnetic fields applied along the *c*-axis. A pronounced anomaly near 119 K, associated with the reduction of spin disorder, corresponds to the reported antiferromagnetic transition[24,28,30]. No anomalies are observed up to 400 K (see **Figure S1**). The resistance increases by more than three orders of magnitude upon cooling from 300 K, indicative of gap-like behavior. The top-left inset displays a fit to the thermally activated conduction model,

$$R(T) = R_0 \exp\left(\frac{E_g}{2k_B T}\right).$$

where $R$ is the resistance at temperature $T$, $R_0$ is a constant related to sample geometry and carrier mobility, $E_g$ is the band-gap energy, and $k_B$ is the Boltzmann constant ($8.617 \times 10^{-5}$ eV/K). The fit yields an energy gap of ~400 meV, confirming the semiconducting nature of YbMn$_2$Sb$_2$. This underlines the importance of sample purity, since doping can substantially modify its electronic properties. The impurity can easily tune the band structure and close the band gap, leading to metallic behavior, as shown in the polycrystalline sample[28]. The top-right inset shows the derivative of resistance, and the bottom-right inset demonstrates that the transition temperature remains nearly unchanged with applied fields up to 90 kOe. The criteria used to determine the transition temperature are provided in **Figure S2**. Additional magnetization measurements and the corresponding phase diagram are shown in the **Figures. S3-S5**. The resistance curves obtained under different field protocols, zero-field-cooled warming (ZFCW) and field cooling (FC), overlap above the transition. **Figures 4b-4d** present the temperature-dependent resistance of YbMn$_2$Sb$_2$ under pressures up to 50.3 GPa, with two independent measurement runs yielding consistent results. Upon application of pressure, the system evolves from semiconducting to metallic behavior. At 1.0 and 2.5 GPa (**Figure 4b**), the resistance is already metallic in character, although a weak low-temperature upturn persists, which vanishes by 3.7 GPa. Below 5.2 GPa, the absolute resistance varies strongly with pressure. In the intermediate range (5.2-19.5 GPa; **Figure 4c**), the metallic behavior is retained, with resistance showing a weaker dependence on pressure. Above 19.5 GPa (**Figure 4d**), the resistance becomes less pressure independent.

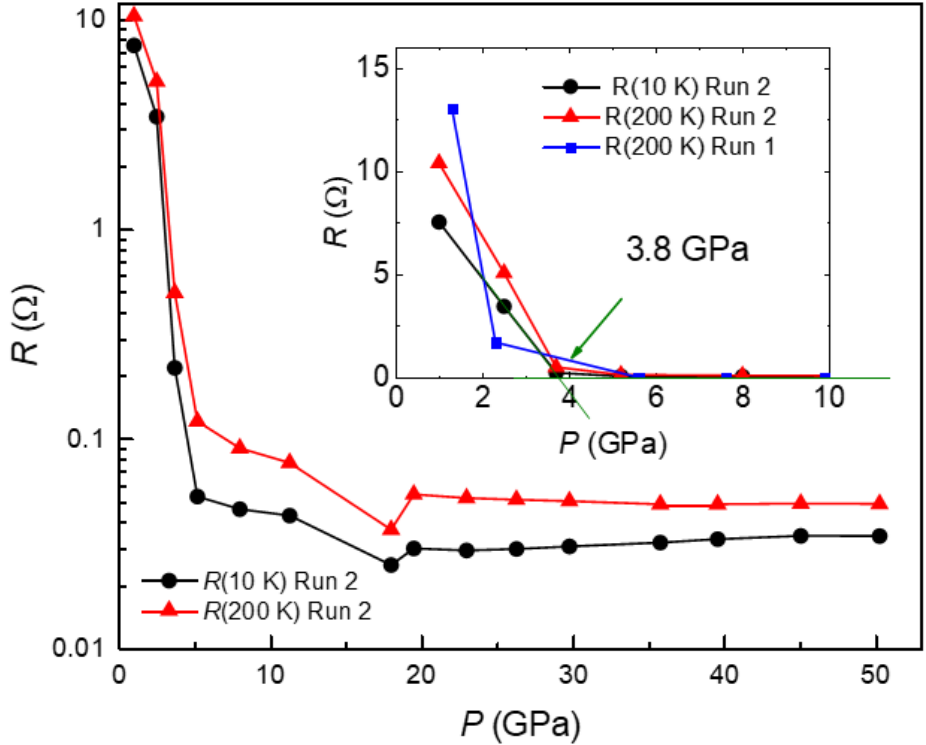

**Figure 5. Pressure-dependent resistance of YbMn$_2$Sb$_2$ measured at 10 K and 200 K using different samples on a log scale.** The inset highlights the crossover pressure separating the regime of dramatic resistance variation from that of minor resistance change.

**Figure 5** summarizes the resistance of YbMn$_2$Sb$_2$ measured at 10 K and 200 K as a function of pressure. As shown in the inset, resistance decreases sharply with pressure below ~4 GPa but becomes nearly pressure independent above this threshold. This behavior is consistent with a structural transition from the trigonal $P\bar{3}m1$ phase to the monoclinic $P2_1/m$ phase near $3.8 \pm 0.5$ GPa. The trigonal $P\bar{3}m1$ phase exhibits semiconducting behavior with an energy gap of ~400 meV, whereas applied pressure modifies the bandwidth and closes the gap. Above ~4 GPa, the monoclinic $P2_1/m$ phase is metallic, with resistance remaining essentially constant under further compression. Due to the sensitivity limits of high-pressure transport measurements, there are no features detected near 119 K in these experiments.

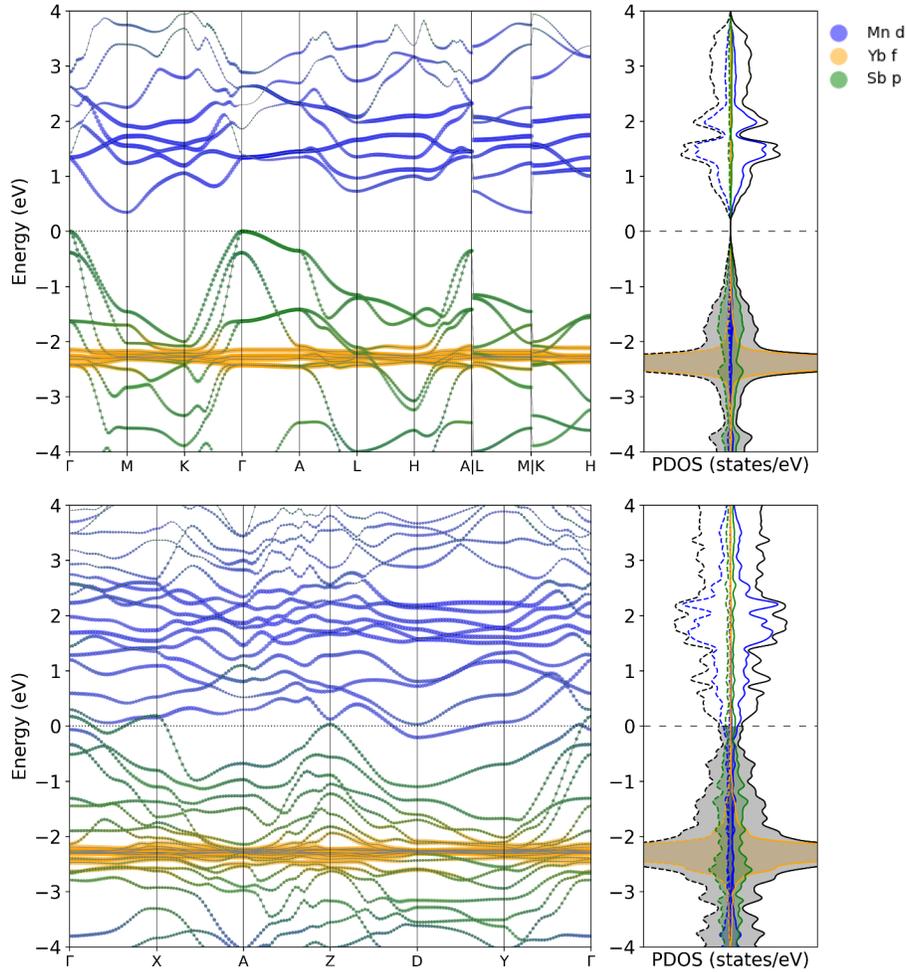

**Figure 6.** The band structures and DOS of YbMn$_2$Sb$_2$ under (***a***) ambient pressure and (***b***) 5.6 GPa.

To corroborate the pressure-induced semiconductor-metal transition, the electronic structures of YbMn$_2$Sb$_2$ were calculated at both ambient and high pressure using first-principles density functional theory. **Figure 6** presents the calculated band structures and projected density of states (PDOS) at ambient pressure (top) and at 5.6 GPa (bottom). At ambient conditions, YbMn$_2$Sb$_2$ exhibits a band gap of ~0.35 eV and a Mn moment of ~4.4 $\mu_B$ per atom, in good agreement with experimental observations. Under compression to 5.6 GPa, the crystal structure evolves such that the gap closes, yielding a metallic state. In this regime, the Yb 4f bands remain well below the Fermi level, while the electronic states near $E$F are dominated by hybridized Sb-*p* and Mn-*d* orbitals. Several magnetic configurations were compared at ambient pressure. The total energies for the nonmagnetic, ferromagnetic, and antiferromagnetic states were calculated to be 0,

–0.23, and –0.25 Ry, respectively, indicating that the AFM state is the most stable, consistent with experiment.

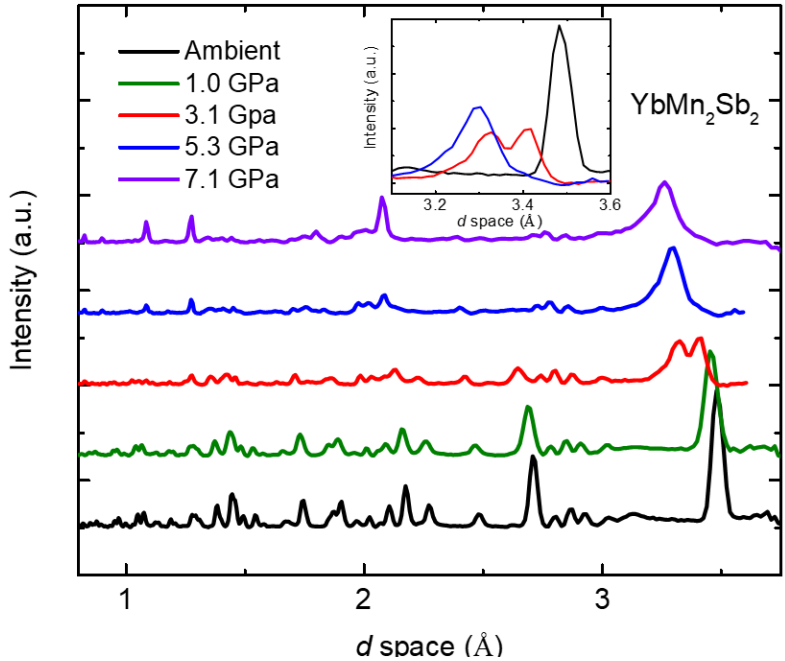

**Figure 7. Neutron powder diffraction (NPD) patterns of YbMn$_2$Sb$_2$ collected at room temperature under various pressures.** The inset highlights data at selected pressures: ambient (before the transition), 3.1 GPa (during the transition), and 5.3 GPa (after the transition).

The monoclinic $P2_1/m$ phase first emerges at ~3.5 GPa, and the coexistence of the two phases persists over an interval of approximately 1 GPa. Within this range, the fraction of the high-pressure monoclinic phase gradually increases until it becomes the sole stable phase. This coexistence is illustrated in **Figure 7**. The two-phase region may originate from pressure inhomogeneity within the sample or from the large volume collapse associated with the transition, which introduces a significant kinetic barrier to completing the transformation uniformly. In the high-pressure regime, the strong diffraction peak near 3.2 Å (**Figure 7**) exhibits substantial broadening. The origin of this feature-whether arising from pressure gradients, sluggish transition kinetics, or intrinsic structural disorder-remains unresolved. Rietveld refinements of the nuclear reflections in the high-pressure phase do not fully converge due to this broadening, representing a major limitation of the fits. In contrast, magnetic reflections collected at the 50° detector bank (lower resolution) are not affected by this issue. Pressure values determined from the Pb calibrant remain consistent with those obtained from the Birch-Murnaghan fits.

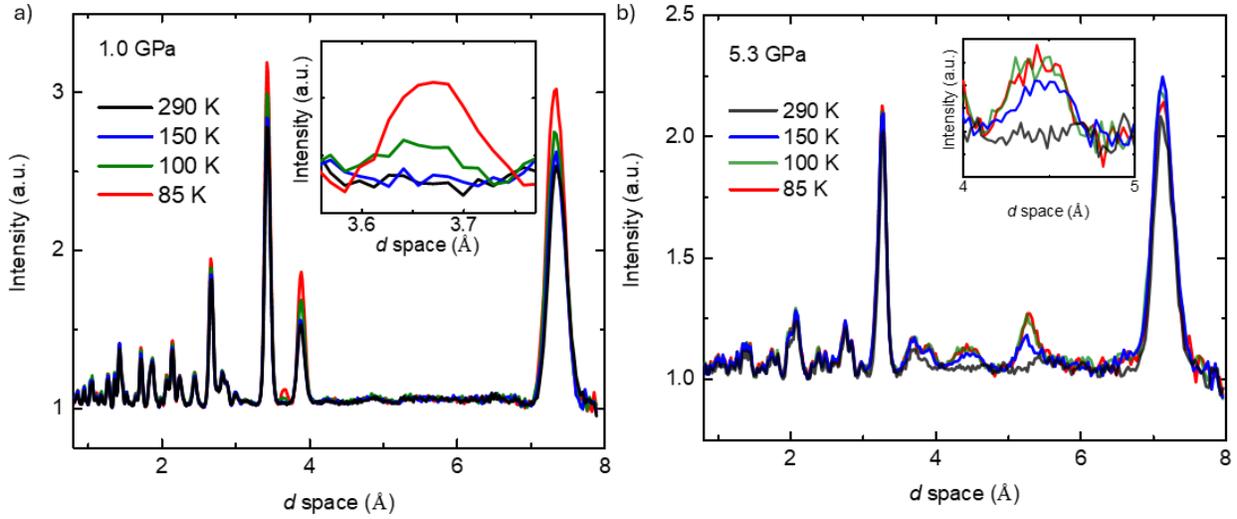

**Figure 8.** Neutron powder diffraction (NPD) patterns of YbMn$_2$Sb$_2$ collected at different temperatures under (*a*) 1.0 GPa and (*b*) 5.3 GPa.

At low pressures (1GPa), shown in **Figure 8a**, magnetic peaks are observed at temperatures as high as 100 K for YbMn$_2$Sb$_2$. These observations agree with prior neutron and magnetic susceptibility measurements. The peaks correspond to a commensurate lattice with a propagation vector $k$ = (0, 0, 0), matching previously reported magnetic structures: YbMn$_2$Sb$_2$ orders antiferromagnetically with the moments tilting from the *ab* plane toward the *c* axis. This structure is described by the magnetic space group $P\bar{1}'$. This magnetic structure persists even upon pressure increasing to 1 GPa. After the structural transition, shown in **Figure 8b**, a different set of magnetic peaks appears, at temperatures below 150 K. A determination of the new propagation vector, based on Gaussian fits of the peaks observed at 4.5 Å and 5.0 Å (with the 4.5 Å peak showing a clear doublet), resulted in an incommensurate propagation vector. Initial best fits yielded $k$ = (0.75, 0.37, 0.25) for YbMn$_2$Sb$_2$. However, relaxing the refinement along all three directions resulted in an improved result with a match with $k$ = (0.742 (6), 0.369 (3), 0.245 (5)) with both the a and c-axis close to the original commensurate values. This structure appears to be temperature dependent upon heating from 85 K to 100 K; the propagation vector slightly changes to $k$ = (0.778 (6), 0.355 (3), 0.267 (5)). The magnetic structure, as discussed further, between 85 K and 100 K is similar, which means the shifting k-vectors could be a result of small changes in anisotropy and exchange interactions. This is similar to what is observed in BaCuSi$_2$O$_6$, where upon cooling, the propagation vector changes due to shifting SiO$_4$ tetrahedra[56]. It should be noted that the magnetic peaks are observed at a low-angle detector (i.e., poor resolution) and are further affected by deviatory stresses on the sample.

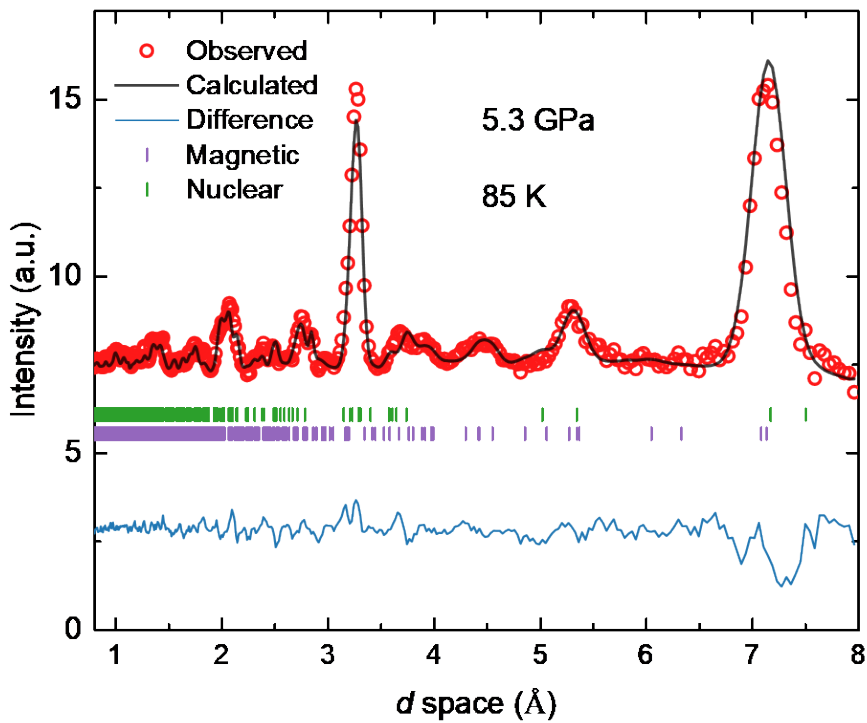

**Figure 9.** Rietveld refinement of neutron powder diffraction (NPD) data for YbMn$_2$Sb$_2$ collected at ~5.3 GPa.

**Figure 9** shows the refinement of the neutron powder diffraction data at 5.3 GPa. Representational analysis using the refined propagation vectors yielded four irreducible representations, each containing three basis vectors (**Table S1**). In the high-pressure monoclinic phase, the two crystallographically distinct Mn sites (Mn1 and Mn2) split into four independent magnetic substructures: Mn$_{1a}$ = ($x$, 0.25, $z$), Mn$_{1b}$ = ($1 - x$, 0.75, $1 - z$), Mn$_{2a}$ = ($u$, 0.25, $w$), and Mn$_{2b}$ = ($1 - u$, 0.75, $1 - w$). Refinement of all mixing coefficients results in a sinusoidal-like magnetic structure reminiscent of that reported for CaMn$_2$Bi$_2$[22] as the Mn pairs (Mn$_{1a}$:Mn$_{2a}$ and Mn$_{1b}$:Mn$_{2b}$) align antiparallel, each carrying a moment of ~4 $\mu_B$ that exhibits slight temperature dependence. While the parent structure features a two-dimensional honeycomb Mn layer, compression drives a transformation into a chain along the $b$-axis. The best-fit magnetic structure involves pairing nearest-neighbor Mn$_{1a}$:Mn$_{2a}$ and Mn$_{1b}$:Mn$_{2b}$ within the $ac$ plane. However, next-nearest ($d_2$) and the third nearest ($d_3$) interactions cannot be neglected, as their interatomic distances are comparable to that of the nearest neighbors: $d_1$ = 2.994(3) Å, $d_2$ = 3.062(3) Å, and $d_3$ = 3.086(3)Å, all confined within the same quasi-1D chain (see **Figure 10**). Interchain coupling, mediated through direct overlap or super-exchange, further contributes to the overall connectivity. The combination of strong exchange interactions along the chains and additional interchain pathways is the likely driver of the observed complex magnetic network in the high-pressure phase.

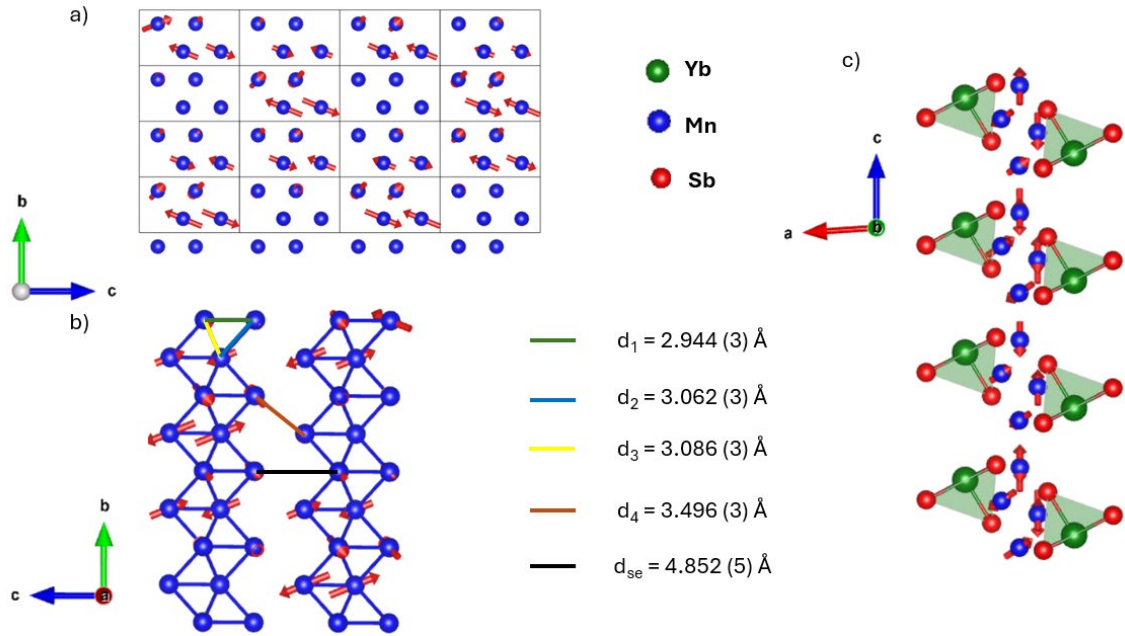

**Figure 10. Magnetic structure of YbMn$_2$Sb$_2$ under pressure.** (*a*) magnetic structure highlighting the quasi-1D chains along the b-axis. (*b*) Magnetic structure of YbMn$_2$Sb$_2$ showing various bond lengths and potential exchange pathways. Note that d$_{se}$ is a potential superexchange pathway between two chains. (*c*) Magnetic structure looking down the *b*-axis. Bonds show the AFM coupling between two Mn atoms within the same chain.

The complex magnetic structure, driven by the incommensurate characters, consists of a sinusoidal propagation of the moment along all 3 lattice directions, where the Mn moment oscillates between a maximum value of ~4.8 μ$_B$ (see **Table 3**) and a minimum of ~0.12 μ$_B$. However, that only remains true for the magnetic structure below 100 K. When the temperature increases to 150 K, the magnetic structure changes as both AFM coupled pairs have a magnetic moment around ~3.7 μ$_B$, a notable reduction for Mn$_{1a}$:Mn$_{2a}$ pairs, but the Mn$_{1a}$:Mn$_{2a}$ retains a similar magnetic moment. As shown in **Figure 8,** as temperature increases, the only peak that is reduced is the peak around 5.3 Å, which corresponds to (-1, 0 1), (0, 1, 0), and (1, 1, 1) magnetic vectors. The high-pressure phase results in magnetic ordering at a higher temperature, which is likely driven by a combination of stronger exchange interactions, new exchange pathways, and reduction of geometric frustration.

**Table 3.** Total and projected components of the magnetic moments of Mn at different temperatures.

| Translation | Crystal axis | 85 K | 100 K | 150 K |
|---|---|---|---|---|
| **Mn$_{1a}$** | | | | |
| | a | 0 | 0 | 0 |
| | b | 2.7 (3) | 2.7 (3) | 3.2 (3) |
| | c | -3.7 (3) | -3.9 (3) | -2.0 (3) |
| **Total Moment** | | 4.6 (3) | 4.7 (3) | 3.8 (3) |
| **Mn$_{1b}$** | | | | |
| | a | 2.2 (3) | 2.1 (4) | 1.7 (4) |
| | b | -2.4 (3) | -2.4 (3) | -1.8 (3) |
| | c | -1.3 (3) | -1.5 (4) | -2.4 (3) |
| **Total Moment** | | 3.6 (3) | 3.6 (4) | 3.5 (3) |
| **Mn$_{2a}$** | | | | |
| | a | 0 | 0 | 0 |
| | b | -2.7 (3) | -2.7 (3) | -3.2 (3) |
| | c | 3.7 (2) | 3.9 (3) | 2.0 (3) |
| **Total Moment** | | 4.6 (3) | 4.7 (3) | 3.8 (3) |
| **Mn$_{2b}$** | | | | |
| | a | -2.2 (3) | -2.1 (4) | -1.7 (4) |
| | b | 2.4 (3) | 2.4 (3) | 1.8 (3) |
| | c | 1.3 (3) | 1.5 (4) | 2.4 (3) |
| **Total Moment** | | 3.6 (3) | 3.6 (4) | 3.5 (3) |

## Conclusion

Single-crystal YbMn$_2$Sb$_2$ was successfully synthesized and comprehensively characterized under ambient and high-pressure conditions. At ambient pressure, structural analysis confirms the reported trigonal $P\bar{3}m1$ phase consisting of alternating Yb, Mn, and Sb layers stacked along the *c*-axis. Temperature-dependent resistance measurements identify semiconducting behavior with a pronounced anomaly at the antiferromagnetic transition temperature ($T_N \approx 119$ K), consistent with long-range magnetic order. Upon compression, YbMn$_2$Sb$_2$ undergoes a semiconductor-to-metal crossover, accompanied by a sharp resistance change near 3-4 GPa. This transition correlates with a structural transformation from the trigonal $P\bar{3}m1$ to the monoclinic $P2_1/m$ phase, as revealed by high-pressure X-ray and neutron diffraction. The transition is accompanied by a substantial volume collapse of about 12% and a phase coexistence over about 1 GPa. In the high-pressure regime (>5 GPa), the material exhibits metallic transport behavior, with diminishing pressure sensitivity; above ~20 GPa, resistance becomes nearly pressure-independent. Neutron diffraction at ambient pressure confirms antiferromagnetic ordering below 119 K, with Mn moments oriented primarily within the *ab*-plane but gradually tilting toward the *c*-axis with pressure. Under the high-pressure monoclinic phase, a different magnetic propagation vector emerges, resulting in a sinusoidal-like arrangement of Mn moments. This quasi-one-dimensional magnetic configuration resembles that observed in the Bi analogue $CaMn_2Bi_2$, although with distinct propagation vectors and moment orientations. These results underscore how modifications of the ligand environment around Mn profoundly influence magnetic structure and demonstrate the critical role of pressure in tuning the coupled electronic, structural, and magnetic properties of $YbMn_2Sb_2$.

## Supporting Information

Supporting Information is available from internet or from the author.

## Acknowledgments

M.X. M.B. C.P. and W.X. at Michigan State University were supported by the U.S. Department of Energy (DOE), Division of Basic Energy Sciences under Contract DE-SC0023648. A. R. and H.D.Z. thank the support from NSF-DMR-2003117. Y.Z. was supported by National Natural Science Foundation of China (12204004). S.W. and Z. Y. were supported by the National Key Research and Development Program of China (Grant No. 2023YFA1406102). A portion of this research used resources at the Spallation Neutron Source and the Center for Nanophase Materials Science both DOE Office of Science User Facility operated by the Oak Ridge National Laboratory. The neutron beam time was allocated to SNAP under proposal IPTS-34311.

## Conflict of Interest

# References

(1) Keimer, B.; Kivelson, S. A.; Norman, M. R.; Uchida, S.; Zaanen, J. From Quantum Matter to High-Temperature Superconductivity in Copper Oxides. *Nature* 2015, *518* (7538), 179–186. https://doi.org/10.1038/nature14165.

(2) Basov, D. N.; Averitt, R. D.; Hsieh, D. Towards Properties on Demand in Quantum Materials. *Nat Mater* 2017, *16* (11), 1077–1088. https://doi.org/10.1038/nmat5017.

(3) Tokura, Y.; Kawasaki, M.; Nagaosa, N. Emergent Functions of Quantum Materials. *Nat Phys* 2017, *13* (11), 1056–1068. https://doi.org/10.1038/nphys4274.

(4) Szytuła, A.; Szott, I. Magnetic Properties of Ternary $RMn_2Si_2$ and $RMn_2Ge_2$ Compounds. *Solid State Commun* 1981, *40* (2), 199–202. https://doi.org/10.1016/0038-1098(81)90167-8.

(5) Simonson, J. W.; Smith, G. J.; Post, K.; Pezzoli, M.; Kistner-Morris, J. J.; McNally, D. E.; Hassinger, J. E.; Nelson, C. S.; Kotliar, G.; Basov, D. N.; Aronson, M. C. Magnetic and Structural Phase Diagram of $CaMn_2Sb_2$. *Phys Rev B* 2012, *86* (18), 184430. https://doi.org/10.1103/PhysRevB.86.184430.

(6) Anand, V. K.; Johnston, D. C. Metallic Behavior Induced by Potassium Doping of the Trigonal Antiferromagnetic Insulator $EuMn_2As_2$. *Phys Rev B* 2016, *94* (1), 014431. https://doi.org/10.1103/PhysRevB.94.014431.

(7) Schmidt, J.; Sapkota, A.; Mueller, C. L.; Xiao, S.; Huyan, S.; Slade, T. J.; Ribeiro, R. A.; Lee, S.-W.; Bud'ko, S. L.; Canfield, P. C. Tuning the Structure and Superconductivity of $SrNi_2P_2$ by Rh Substitution. *Phys Rev B* 2025, *111* (5), 054102. https://doi.org/10.1103/PhysRevB.111.054102.

(8) Jo, N. H.; Kuthanazhi, B.; Wu, Y.; Timmons, E.; Kim, T.-H.; Zhou, L.; Wang, L.-L.; Ueland, B. G.; Palasyuk, A.; Ryan, D. H.; McQueeney, R. J.; Lee, K.; Schrunk, B.; Burkov, A. A.; Prozorov, R.; Bud'ko, S. L.; Kaminski, A.; Canfield, P. C. Manipulating Magnetism in the Topological Semimetal $EuCd_2As_2$. *Phys Rev B* 2020, *101* (14), 140402. https://doi.org/10.1103/PhysRevB.101.140402.

(9) Ruszała, P.; Winiarski, M. J.; Samsel-Czekała, M.; Tran, L. M.; Babij, M.; Bukowski, Z. Tilted Dirac Cone Gapped Due to Spin-Orbit Coupling and Transport Properties of a 3D Metallic System $CaIr_2Ge_2$. *J Alloys Compd* 2021, *858*, 158260. https://doi.org/10.1016/j.jallcom.2020.158260.

(10) Yang, X. P.; Yao, Y.-T.; Zheng, P.; Guan, S.; Zhou, H.; Cochran, T. A.; Lin, C.-M.; Yin, J.-X.; Zhou, X.; Cheng, Z.-J.; Li, Z.; Shi, T.; Hossain, M. S.; Chi, S.; Belopolski, I.; Jiang, Y.-X.; Litskevich, M.; Xu, G.; Tian, Z.; Bansil, A.; Yin, Z.; Jia, S.; Chang, T.-R.; Hasan, M. Z. A Topological Hund Nodal Line Antiferromagnet. *Nat Commun* 2024, *15* (1), 7052. https://doi.org/10.1038/s41467-024-51255-3.

(11) Gascoin, F.; Ottensmann, S.; Stark, D.; Haïle, S. M.; Snyder, G. J. Zintl Phases as Thermoelectric Materials: Tuned Transport Properties of the Compounds $Ca_xYb_{1-x}Zn_2Sb_2$. *Adv Funct Mater* 2005, *15* (11), 1860–1864. https://doi.org/10.1002/adfm.200500043.

(12) Zhang, H.; Zhao, J.-T.; Grin, Yu.; Wang, X.-J.; Tang, M.-B.; Man, Z.-Y.; Chen, H.-H.; Yang, X.-X. A New Type of Thermoelectric Material, $EuZn_2Sb_2$. *J Chem Phys* 2008, *129* (16). https://doi.org/10.1063/1.3001608.

(13) Mao, J.; Zhu, H.; Ding, Z.; Liu, Z.; Gamage, G. A.; Chen, G.; Ren, Z. High Thermoelectric Cooling Performance of N-Type $Mg_3Bi_2$-Based Materials. *Science (1979)* 2019, *365* (6452), 495–498. https://doi.org/10.1126/science.aax7792.

(14) Fukazawa, H.; Takeshita, N.; Yamazaki, T.; Kondo, K.; Hirayama, K.; Kohori, Y.; Miyazawa, K.; Kito, H.; Eisaki, H.; Iyo, A. Suppression of Magnetic Order by Pressure in $BaFe_2As_2$. *J Physical Soc Japan* 2008, *77* (10), 105004. https://doi.org/10.1143/JPSJ.77.105004.

(15) Sefat, A. S.; Jin, R.; McGuire, M. A.; Sales, B. C.; Singh, D. J.; Mandrus, D. Superconductivity at 22 K in Co-Doped $BaFe_2As_2$ Crystals. *Phys Rev Lett* 2008, *101* (11), 117004. https://doi.org/10.1103/PhysRevLett.101.117004.

(16) Iyo, A.; Kawashima, K.; Kinjo, T.; Nishio, T.; Ishida, S.; Fujihisa, H.; Gotoh, Y.; Kihou, K.; Eisaki, H.; Yoshida, Y. New-Structure-Type Fe-Based Superconductors: $CaAFe_4As_4$ (A =K, Rb, Cs) and $SrAFe_4As_4$ (A = Rb, Cs). *J Am Chem Soc* 2016, *138* (10), 3410–3415. https://doi.org/10.1021/jacs.5b12571.

(17) Xu, M.; Schmidt, J.; Gati, E.; Xiang, L.; Meier, W. R.; Kogan, V. G.; Bud'ko, S. L.; Canfield, P. C. Superconductivity and Phase Diagrams of $CaK(Fe_{1-x}Mn_x)_4As_4$ Single Crystals. *Phys Rev B* 2022, *105* (21), 214526. https://doi.org/10.1103/PhysRevB.105.214526.

(18) Sangeetha, N. S.; Pakhira, S.; Ding, Q.-P.; Krause, L.; Lee, H.-C.; Smetana, V.; Mudring, A.-V.; Iversen, B. B.; Furukawa, Y.; Johnston, D. C. First-Order Antiferromagnetic Transitions of $SrMn_2P_2$ and $CaMn_2P_2$ Single Crystals Containing Corrugated-Honeycomb Mn Sublattices. *Proceedings of the National Academy of Sciences* 2021, *118* (44). https://doi.org/10.1073/pnas.2108724118.

(19) Chang, T.; Pletikosic, I.; Kong, T.; Bian, G.; Huang, A.; Denlinger, J.; Kushwaha, S. K.; Sinkovic, B.; Jeng, H.; Valla, T.; Xie, W.; Cava, R. J. Realization of a Type-II Nodal-Line Semimetal in $Mg_3Bi_2$. *Advanced Science* 2019, *6* (4). https://doi.org/10.1002/advs.201800897.

(20) Gibson, Q. D.; Wu, H.; Liang, T.; Ali, M. N.; Ong, N. P.; Huang, Q.; Cava, R. J. Magnetic and Electronic Properties of CaMn$_2$Bi$_2$: A Possible Hybridization Gap Semiconductor. *Phys Rev B* 2015, *91* (8), 085128. https://doi.org/10.1103/PhysRevB.91.085128.

(21) Gui, X.; Finkelstein, G. J.; Chen, K.; Yong, T.; Dera, P.; Cheng, J.; Xie, W. Pressure-Induced Large Volume Collapse, Plane-to-Chain, Insulator to Metal Transition in CaMn$_2$Bi$_2$. *Inorg Chem* 2019, *58* (14), 8933–8937. https://doi.org/10.1021/acs.inorgchem.9b01362.

(22) Marshall, M.; Wang, H.; dos Santos, A. M.; Haberl, B.; Xie, W. Incommensurate Spiral Spin Order in CaMn$_2$Bi$_2$ Observed via High-Pressure Neutron Diffraction. *Inorg Chem* 2024, *63* (4), 1736–1744. https://doi.org/10.1021/acs.inorgchem.3c02379.

(23) Rühl, R.; Jeitschko, W. New Pnictides with Ce$_2$O$_2$S-Type Structure. *Mater Res Bull* 1979, *14* (4), 513–517. https://doi.org/10.1016/0025-5408(79)90194-6.

(24) Morozkin, A. V.; Isnard, O.; Henry, P.; Granovsky, S.; Nirmala, R.; Manfrinetti, P. Synthesis and Magnetic Structure of the YbMn$_2$Sb$_2$ Compound. *J Alloys Compd* 2006, *420* (1–2), 34–36. https://doi.org/10.1016/j.jallcom.2005.10.051.

(25) Liu, K.; Liu, J.; Liu, Q.; Wang, Q.; Yu, F.; Liu, X.-C.; Xia, S.-Q. Phase Transitions, Structure Evolution, and Thermoelectric Properties Based on A$_2$MnSb$_2$ (A = Ca, Yb). *Chemistry of Materials* 2021, *33* (24), 9732–9740. https://doi.org/10.1021/acs.chemmater.1c03527.

(26) Nikiforov, V. N.; Pryadun, V. V.; Morozkin, A. V.; Irkhin, V. Yu. Anomalies of Transport Properties in Antiferromagnetic YbMn$_2$Sb$_2$ Compound. *Phys Lett A* 2014, *378* (20), 1425–1427. https://doi.org/10.1016/j.physleta.2014.03.030.

(27) Liu, S.; Li, A.; Hao, X.; Chen, M.; Huang, Y.; Yu, Y.; Wang, C.-W.; Ren, Q.; Zhu, T.; Fu, C. Synthesis and Thermal Stability of Topological Semimetal *R*MnSb$_2$ ( *R* = Yb, Sr, Ba, Eu). *Sci Technol Adv Mater* 2025, *26* (1). https://doi.org/10.1080/14686996.2025.2512702.

(28) Nirmala, R.; Morozkin, A. V.; Suresh, K. G.; Kim, H.-D.; Kim, J.-Y.; Park, B.-G.; Oh, S.-J.; Malik, S. K. Magnetism and Transport in YbMn$_2$Sb$_2$. *J Appl Phys* 2005, *97* (10). https://doi.org/10.1063/1.1851815.

(29) Engel, S.; Gießelmann, E. C. J.; Reimann, M. K.; Pöttgen, R.; Janka, O. On the Ytterbium Valence and the Physical Properties in Selected Intermetallic Phases. *ACS Organic & Inorganic Au* 2024, *4* (2), 188–222. https://doi.org/10.1021/acsorginorgau.3c00054.

(30) Munevar, J.; Arantes, F. R.; Mendonça-Ferreira, L.; Avila, M. A.; Ribeiro, R. A. Study of the Magnetic Properties of YbMn$_2$Sb$_2$ Single Crystals by MSR. *J Magn Magn Mater* 2021, *537*, 168149. https://doi.org/10.1016/j.jmmm.2021.168149.

(31) Canfield, P. C. New Materials Physics. *Reports on Progress in Physics* 2020, *83* (1), 016501. https://doi.org/10.1088/1361-6633/ab514b.

(32) Canfield, P. C.; Kong, T.; Kaluarachchi, U. S.; Jo, N. H. Use of Frit-Disc Crucibles for Routine and Exploratory Solution Growth of Single Crystalline Samples. *Philosophical Magazine* 2016, *96* (1), 84–92. https://doi.org/10.1080/14786435.2015.1122248.

(33) LSP INDUSTRIAL CERAMICS. Canfield Crucible Sets. 2015.

(34) Parkin, S.; Moezzi, B.; Hope, H. XABS 2: An Empirical Absorption Correction Program. *J Appl Crystallogr* 1995, *28* (1), 53–56. https://doi.org/10.1107/S0021889894009428.

(35) Walker, N.; Stuart, D. An Empirical Method for Correcting Diffractometer Data for Absorption Effects. *Acta Crystallogr A* 1983, *39* (1), 158–166. https://doi.org/10.1107/S0108767383000252.

(36) Sheldrick, G. M. SHELXT – Integrated Space-Group and Crystal-Structure Determination. *Acta Crystallogr A Found Adv* 2015, *71* (1), 3–8. https://doi.org/10.1107/S2053273314026370.

(37) Sheldrick, G. M. Crystal Structure Refinement with SHELXL. *Acta Crystallogr C Struct Chem* 2015, *71* (1), 3–8. https://doi.org/10.1107/S2053229614024218.

(38) Chen, X.; Lou, H.; Zeng, Z.; Cheng, B.; Zhang, X.; Liu, Y.; Xu, D.; Yang, K.; Zeng, Q. Structural Transitions of 4:1 Methanol–Ethanol Mixture and Silicone Oil under High Pressure. *Matter and Radiation at Extremes* 2021, *6* (3). https://doi.org/10.1063/5.0044893.

(39) Dewaele, A.; Torrent, M.; Loubeyre, P.; Mezouar, M. Compression Curves of Transition Metals in the Mbar Range: Experiments and Projector Augmented-Wave Calculations. *Phys Rev B* 2008, *78* (10), 104102. https://doi.org/10.1103/PhysRevB.78.104102.

(40) Mao, H. K.; Xu, J.; Bell, P. M. Calibration of the Ruby Pressure Gauge to 800 Kbar under Quasi-hydrostatic Conditions. *J Geophys Res Solid Earth* 1986, *91* (B5), 4673–4676. https://doi.org/10.1029/JB091iB05p04673.

(41) Piermarini, G. J.; Block, S.; Barnett, J. D.; Forman, R. A. Calibration of the Pressure Dependence of the R1 Ruby Fluorescence Line to 195 Kbar. *J Appl Phys* 1975, *46* (6), 2774–2780. https://doi.org/10.1063/1.321957.

(42) Shen, G.; Wang, Y.; Dewaele, A.; Wu, C.; Fratanduono, D. E.; Eggert, J.; Klotz, S.; Dziubek, K. F.; Loubeyre, P.; Fat'yanov, O. V.; Asimow, P. D.; Mashimo, T.; Wentzcovitch, R. M. M. Toward an International Practical Pressure Scale: A Proposal for an IPPS Ruby Gauge (IPPS-Ruby2020). *High Press Res* 2020, *40* (3), 299–314. https://doi.org/10.1080/08957959.2020.1791107.

(43) Giannozzi, P.; Andreussi, O.; Brumme, T.; Bunau, O.; Buongiorno Nardelli, M.; Calandra, M.; Car, R.; Cavazzoni, C.; Ceresoli, D.; Cococcioni, M.; Colonna, N.; Carnimeo, I.; Dal Corso, A.; de Gironcoli, S.; Delugas, P.; DiStasio, R. A.; Ferretti, A.; Floris, A.; Fratesi, G.; Fugallo, G.; Gebauer, R.; Gerstmann, U.; Giustino, F.; Gorni, T.; Jia, J.; Kawamura, M.; Ko, H.-Y.; Kokalj, A.; Küçükbenli, E.; Lazzeri, M.; Marsili, M.; Marzari, N.; Mauri,

F.; Nguyen, N. L.; Nguyen, H.-V.; Otero-de-la-Roza, A.; Paulatto, L.; Poncé, S.; Rocca, D.; Sabatini, R.; Santra, B.; Schlipf, M.; Seitsonen, A. P.; Smogunov, A.; Timrov, I.; Thonhauser, T.; Umari, P.; Vast, N.; Wu, X.; Baroni, S. Advanced Capabilities for Materials Modelling with Quantum ESPRESSO. *Journal of Physics: Condensed Matter* 2017, *29* (46), 465901. https://doi.org/10.1088/1361-648X/aa8f79.

(44) Giannozzi, P.; Baroni, S.; Bonini, N.; Calandra, M.; Car, R.; Cavazzoni, C.; Ceresoli, D.; Chiarotti, G. L.; Cococcioni, M.; Dabo, I.; Dal Corso, A.; de Gironcoli, S.; Fabris, S.; Fratesi, G.; Gebauer, R.; Gerstmann, U.; Gougoussis, C.; Kokalj, A.; Lazzeri, M.; Martin-Samos, L.; Marzari, N.; Mauri, F.; Mazzarello, R.; Paolini, S.; Pasquarello, A.; Paulatto, L.; Sbraccia, C.; Scandolo, S.; Sclauzero, G.; Seitsonen, A. P.; Smogunov, A.; Umari, P.; Wentzcovitch, R. M. QUANTUM ESPRESSO: A Modular and Open-Source Software Project for Quantum Simulations of Materials. *Journal of Physics: Condensed Matter* 2009, *21* (39), 395502. https://doi.org/10.1088/0953-8984/21/39/395502.

(45) Prandini, G.; Marrazzo, A.; Castelli, I. E.; Mounet, N.; Marzari, N. Precision and Efficiency in Solid-State Pseudopotential Calculations. *NPJ Comput Mater* 2018, *4* (1), 72. https://doi.org/10.1038/s41524-018-0127-2.

(46) Perdew, J. P.; Burke, K.; Ernzerhof, M. Generalized Gradient Approximation Made Simple. *Phys Rev Lett* 1996, *77* (18), 3865–3868. https://doi.org/10.1103/PhysRevLett.77.3865.

(47) Monkhorst, H. J.; Pack, J. D. Special Points for Brillouin-Zone Integrations. *Phys Rev B* 1976, *13* (12), 5188–5192. https://doi.org/10.1103/PhysRevB.13.5188.

(48) Davidson, E. R. The Iterative Calculation of a Few of the Lowest Eigenvalues and Corresponding Eigenvectors of Large Real-Symmetric Matrices. *J Comput Phys* 1975, *17* (1), 87–94. https://doi.org/10.1016/0021-9991(75)90065-0.

(49) Togo, A.; Shinohara, K.; Tanaka, I. Spglib: A Software Library for Crystal Symmetry Search. 2018.

(50) Setyawan, W.; Curtarolo, S. High-Throughput Electronic Band Structure Calculations: Challenges and Tools. *Comput Mater Sci* 2010, *49* (2), 299–312. https://doi.org/10.1016/j.commatsci.2010.05.010.

(51) Bireckoven, B.; Wittig, J. A Diamond Anvil Cell for the Investigation of Superconductivity under Pressures of up to 50 GPa: Pb as a Low Temperature Manometer. *J Phys E* 1988, *21* (9), 841–848. https://doi.org/10.1088/0022-3735/21/9/004.

(52) Toby, B. H.; Von Dreele, R. B. *GSAS-II*: The Genesis of a Modern Open-Source All Purpose Crystallography Software Package. *J Appl Crystallogr* 2013, *46* (2), 544–549. https://doi.org/10.1107/S0021889813003531.

(53) Rodríguez-Carvajal, J. Recent Developments of the Program FULLPROF, in Commission on Powder Diffraction (IUCr). *Newsletter* 2001, *26*, 12–19.

(54) Wills, A. S. A New Protocol for the Determination of Magnetic Structures Using Simulated Annealing and Representational Analysis (SARAh). *Physica B Condens Matter* 2000, *276–278*, 680–681. https://doi.org/10.1016/S0921-4526(99)01722-6.

(55) Fischer, R. A.; Campbell, A. J.; Chidester, B. A.; Reaman, D. M.; Thompson, E. C.; Pigott, J. S.; Prakapenka, V. B.; Smith, J. S. Equations of State and Phase Boundary for Stishovite and CaCl$_2$-Type SiO$_2$. *American Mineralogist* 2018, *103* (5), 792–802. https://doi.org/10.2138/am-2018-6267.

(56) Samulon, E. C.; Islam, Z.; Sebastian, S. E.; Brooks, P. B.; McCourt, M. K.; Ilavsky, J.; Fisher, I. R. Low-Temperature Structural Phase Transition and Incommensurate Lattice Modulation in the Spin-Gap Compound BaCuSi$_2$O$_6$. *Phys Rev B* 2006, *73* (10), 100407. https://doi.org/10.1103/PhysRevB.73.100407.

# Supplementary Information

## Pressure-Induced Structural and Magnetic Evolution in Layered Antiferromagnet YbMn$_2$Sb$_2$

*Mingyu Xu[1], Matt Boswell[1,2], Aya Rutherford[3], Cheng Peng[1], Ying Zhou[4], Shuyang Wang[5], Zhaorong Yang[4,5], Antonio M. dos Santos[2], Haidong Zhou[3], Weiwei Xie[1*]*

1. Department of Chemistry, Michigan State University, East Lansing, MI, 48864, USA
2. Neutron Scattering Division, Oak Ridge National Laboratory, Oak Ridge, TN 37831 USA
3. Department of Physics and Astronomy, University of Tennessee, Knoxville, TN 37996, USA
4. Institutes of Physical Science and Information Technology, Anhui University, Hefei 230601, China
5. Anhui Province Key Laboratory of Low-Energy Quantum Materials and Devices, High Magnetic Field Laboratory, HFIPS, Chinese Academy of Sciences, Hefei, Anhui 230031, China
6.

[*]Corresponding author: Dr. Weiwei Xie (xieweiwe@msu.edu)

## Table of Contents

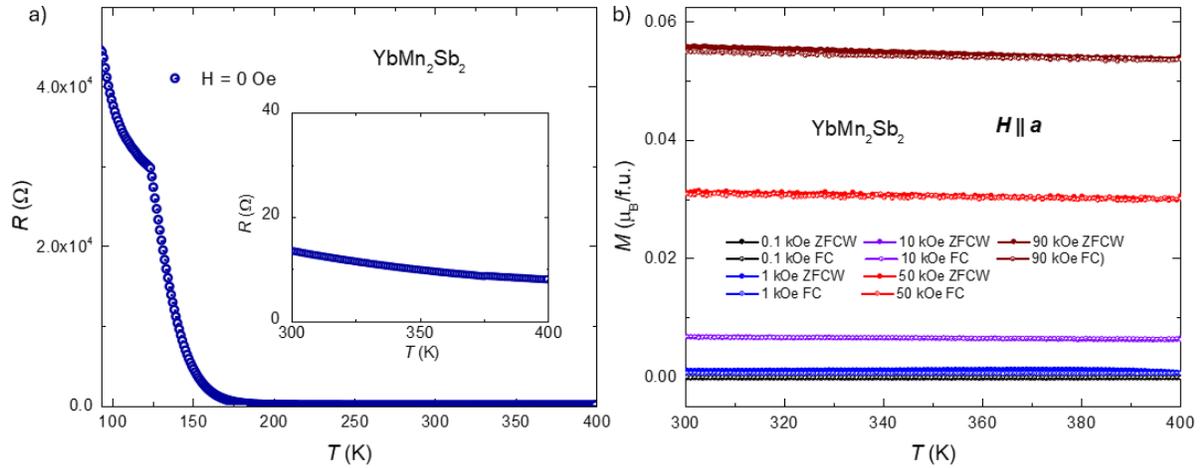

**Figure S1. Resistance and magnetization measurements were conducted up to 400 K.** *a)* Temperature-dependent resistance up to 400 K. The inset gives the measurement in the range from 300 K to 400 K, with a not-observed feature. *b)* Temperature-dependent magnetization at several fields in the temperature range from 300 K to 400 K.

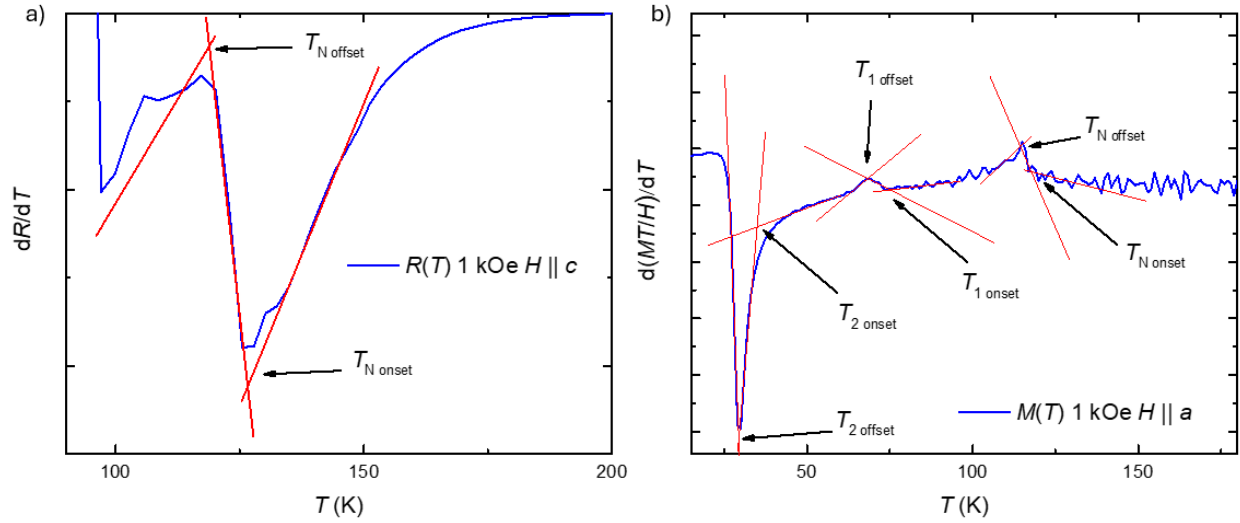

**Figure S2. The criteria of transition temperatures were determined using temperature-dependent resistance and magnetization**. The onset and offset of transition temperatures are shown in **Fig. S3**. The transition temperature is given by the average of the onset and the offset values, and half of the difference between the onset and the offset gives the transition width.

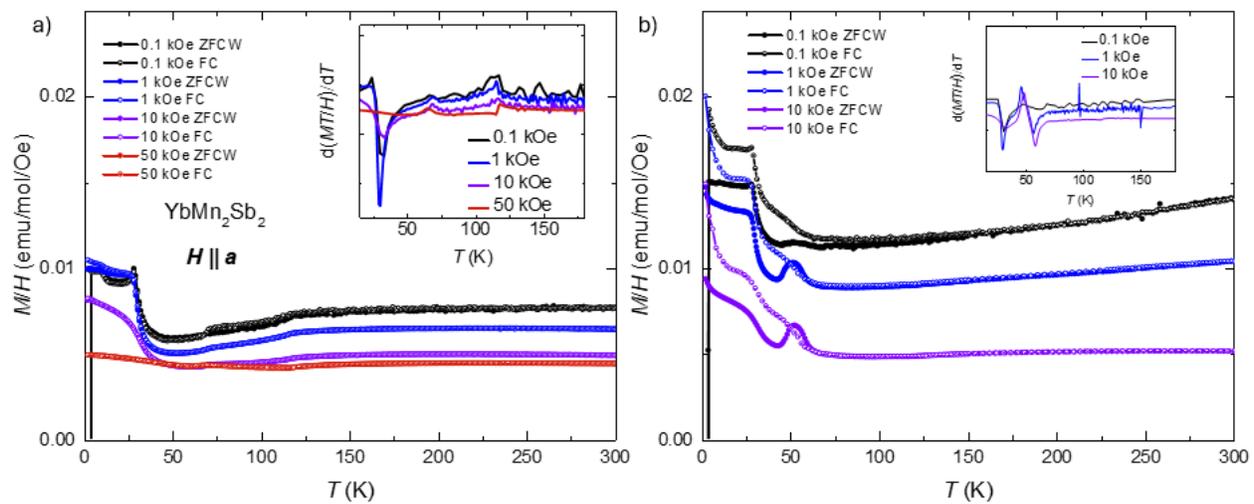

**Figure S3.** *M/H* changes as the temperature changes at different magnetic fields parallel to the *a* and *c* directions (panels a) and b), respectively.

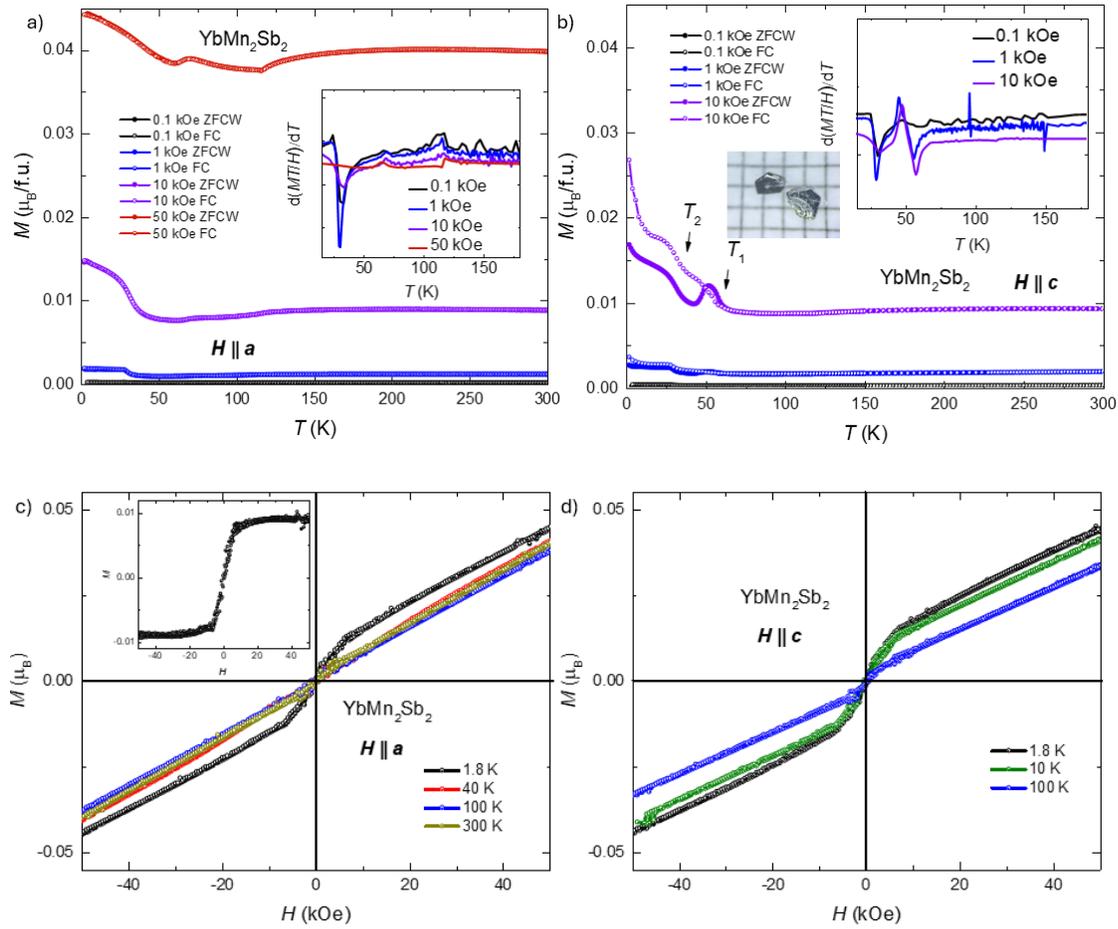

**Figure S4. The anisotropic magnetic moment measurements as a function of temperature and magnetic field**. *a)* Temperature-dependent magnetic moment for different fields along the *a*-axis with two protocols (ZFCW and FC). The inset presents the derivative of the magnetic moment. *b)* Temperature-dependent magnetic moment for different fields along the *c*-axis with two protocols (ZFCW and FC). The left inset gives the picture of synthesized YbMn$_2$Sb$_2$ crystals on millimeter grid paper. The right inset shows the derivative of the magnetic moment. *c)* Field-dependent magnetic moment in the magnetic field along the *a*-axis at different temperatures. The inset shows the *M(H)* without linear behavior. *d)* Field-dependent magnetic moment in the magnetic field along the *c*-axis at different temperatures.

**Fig. S4a** and **S4b** present measurements along *a* and *c* axis of the temperature-dependent magnetic moment under different fields using both ZFCW and FC protocols. An antiferromagnetic transition is shown when the field is applied along the *a*-axis, whereas no detectable anomaly is observed for $H \parallel c$, likely originating from the ~62° tilt of the Mn moments relative to the *c*-axis [20], which introduces a small *c*-axis component to the magnetic moment. When the field is parallel to the *a*-axis, an additional anomaly appears near 40 K, marked as $T_2$, where the moment increases upon cooling. This feature becomes clearer in plots of *M*/*H* versus temperature (**Fig. S3**) and is also shown for $H \parallel c$. Unlike the antiferromagnetic transition, this feature exhibits apparent hysteresis, indicating the emergence of a small net moment. This net moment (~0.01 $\mu_B$/f.u.) first develops around 60 K ($T_1$), is anisotropic at higher temperatures, and becomes isotropic upon cooling to ~40 K. **Fig. S4c** and **S4d** show field-dependent magnetization at selected temperatures. A distinct change in slope is observed below $T_1$ or $T_2$, consistent with the *M*(*T*) data. The insets display the magnetization after subtracting the dominant linear *M*(*H*) background, revealing a net moment increase of ~0.01 $\mu_B$/f.u., in agreement with the temperature-dependent results.

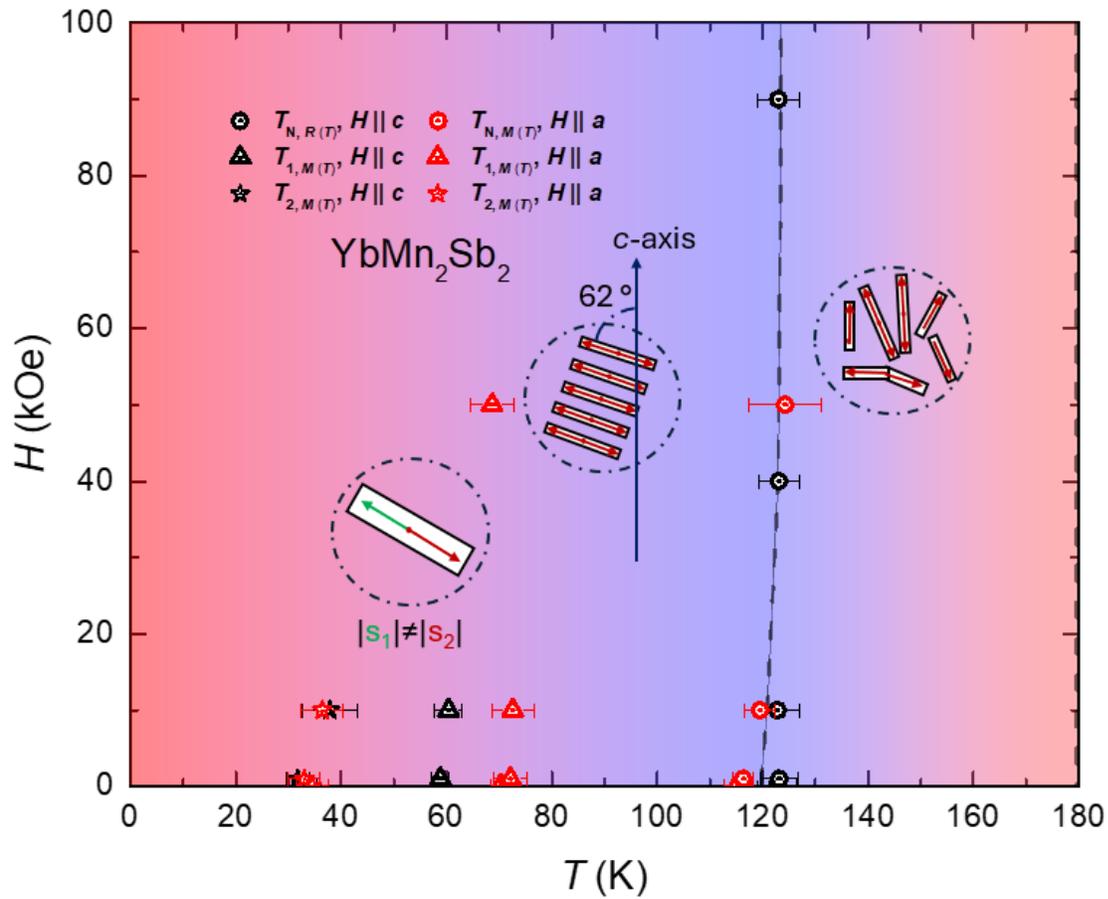

**Figure S5. H-T phase diagram of YbMn$_2$Sb$_2$.**

**Figure S5** shows the *H-T* phase diagram. The blue color indicates that the pair of Mn aligns well, which gives no net moment, and the red color indicates that a small net moment exists. The dashed line gives the long-range order boundary.

**Fig. S5** and **Fig. S2*b*** present the magnetization measurements. Across the whole temperature range from 1.8 K to 400 K, no Curie-Weiss tail is observed in any of the magnetization curves. As shown in **Fig. 3**, YbMn$_2$Sb$_2$ is semiconducting at ambient pressure. It is therefore surprising that neither Mn nor Yb exhibits a bare moment, with the measured magnetization remaining below 0.05 $\mu_B$/f.u., even under an applied field of 50 kOe. In contrast, neutron diffraction experiments reveal an ordered moment of approximately 4 $\mu_B$/f.u. for Mn once long-range magnetic order sets in.

This raises the question: where is the Mn moment above the antiferromagnetic transition in semiconducting YbMn$_2$Sb$_2$? We propose that some of the Mn atoms form pairs of local moments with equal magnitudes but opposite directions, or with moments opposite but do not perfectly align, resulting in the disappearance of the magnetic moment in the absence of long-range order. Upon cooling to 119 K, these moment pairs establish long-range antiferromagnetic order. At lower temperatures, as reported in Ref. [17], the Mn–Mn distance changes significantly. This variation may indicate subtle structural distortions and modify the magnitude of the individual moments. Such changes could generate a small net moment of about 0.01 $\mu_B$ per Mn pair (~0.25%), detectable only against the otherwise very low magnetic background.

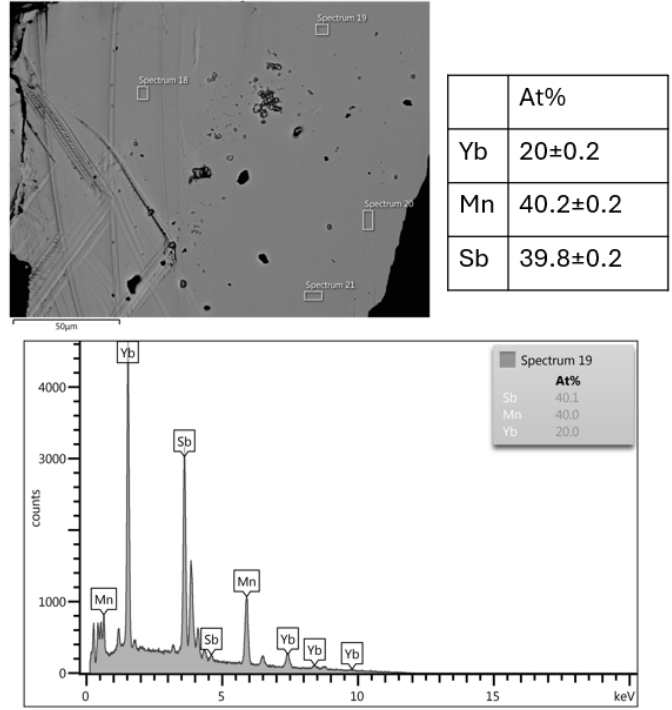

**Figure S6.** EDS results with the unit of atomic percentage.

**Table S1.** Magnetic basis vectors used for the refinement of Mn magnetic structures in YbMn$_2$Sb$_2$. Mn$_1$ and Mn$_2$ are crystallographically distinct sites, each split into two magnetic sublattices (*a* and *b*).

| IR | $\Gamma_1$ | | |
|---|---|---|---|
| BV | M$_a$ | M$_b$ | M$_c$ |
| **Mn$_{1a}$ ψ$_1$** | 1 | 0 | 0 |
| **Mn$_{1a}$ ψ$_2$** | 0 | 1 | 0 |
| **Mn$_{1a}$ ψ$_3$** | 0 | 0 | 1 |
| **Mn$_{1b}$ ψ$_1$** | 1 | 0 | 0 |
| **Mn$_{1b}$ ψ$_2$** | 0 | 1 | 0 |
| **Mn$_{1b}$ ψ$_3$** | 0 | 0 | 1 |
| **Mn$_{2a}$ ψ$_1$** | 1 | 0 | 0 |
| **Mn$_{2a}$ ψ$_2$** | 0 | 1 | 0 |
| **Mn$_{2a}$ ψ$_3$** | 0 | 0 | 1 |
| **Mn$_{2b}$ ψ$_1$** | 1 | 0 | 0 |
| **Mn$_{2b}$ ψ$_2$** | 0 | 1 | 0 |
| **Mn$_{2b}$ ψ$_3$** | 0 | 0 | 1 |